\documentclass[12pt,a4paper]{article}

\usepackage{graphicx}		
\usepackage{lineno}       		
\usepackage{natbib}
\usepackage{mnemonic}  
\usepackage[greek,english]{babel} 

\newcommand{\ignore}[1]{}   

\begin{document}
\sloppy				
\widowpenalty=300
\clubpenalty=300

\parskip 0pt          
\parindent 20pt
\pagestyle{plain}
\pagenumbering{arabic} 

\null
\vspace{20pt}
\centerline{\large \bf The History of Astrometry}
\centerline{Michael Perryman, University of Bristol}
\vspace{20pt}

\centerline{Accepted for publication in EPJ--H, The European Physical Journal}
\centerline{(Historical Perspectives on Contemporary Physics), 28~Aug 2012}
\centerline{The final publication is available at www.epj.org}

\section{Abstract}
The history of astrometry, the branch of astronomy dealing with the positions of celestial objects, is a lengthy and complex chronicle, having its origins in the earliest records of astronomical observations more than two thousand years ago, and extending to the high accuracy observations being made from space today. Improved star positions progressively opened up and advanced fundamental fields of scientific enquiry, including our understanding of the scale of the solar system, the details of the Earth's motion through space, and the comprehension and acceptance of Newtonianism. They also proved crucial to the practical task of maritime navigation. Over the past  400~years, during which positional accuracy has improved roughly logarithmically with time, the distances to the nearest stars were triangulated, making use of the extended measurement baseline given by the Earth's orbit around the Sun. This led to quantifying the extravagantly vast scale of the Universe, to a determination of the physical properties of stars, and to the resulting characterisation of the structure, dynamics and origin of our Galaxy. After a period in the middle years of the twentieth century in which accuracy improvements were greatly hampered by the perturbing effects of the Earth's atmosphere, ultra-high accuracies of star positions from space platforms have led to a renewed advance in this fundamental science over the past few years.

\section{Introduction}

\subsection{The Context}

Astrometry is the branch of astronomy concerned with the accurate measurement of the positions and motions of celestial objects. This includes the positions and motions of the planets and other solar system bodies, stars within our Galaxy and, in principle, galaxies and clusters of galaxies within the Universe. Since recording and refining the positions of the stars and planets on the sky was one of the few investigations of the heavens open to the ancients, astronomy and astrometry were largely synonymous until a little more than a century ago, when other types of astronomical investigation, such as spectroscopy, became possible. 

Astrometry therefore has a remarkably long scientific history, while it remains acutely topical today. Over more than two millennia of recorded history, star positions have been measured with progressively increasing accuracy and, for reasons that will become evident, fundamental advances in our understanding of the Universe have accompanied this progress. As in all sciences, advances have traced out a perpetual contest between theory and observation. New ideas down the centuries demanded better observations to confirm them, while instrumental advances provided new empirical evidence which in turn stimulated new ideas. Throughout this history, advances in astrometry have benefitted from telescope improvements, from the control of measurement errors and, in particular, from the ability to graduate and further subdivide angular arcs on the celestial sphere. Recording star positions progressed through naked eye observations, later assisted by optical telescopes, through to the large-scale recording of stellar images on photographic plates of the late nineteenth century, to the high-efficiency solid-state detectors of the last twenty years.

Some 50--100 years ago, progress ran into almost insurmountable problems imposed by the Earth's atmosphere. Scientific advances associated with improved astrometric measurements faltered, and astrometry consequently receded in global scientific importance in the face of many other rapidly-advancing branches of observational and theoretical astronomy, such as spectroscopy and cosmology. It took a back seat as observations opened up in other electromagnetic frequency domains, such as in the radio, infrared, and X-ray. Within the past two decades, unprecedented positional observations made from satellites placed above the Earth's atmosphere have ushered in a renewed and unparalleled advance in astrometric measurement capability.

The history of astrometry is a large and multiply-connected field. I will lay out a selective summary aiming to pinpoint some relevant highlights as it developed from its earliest roots. I will relate how, over the centuries, improved star positions led to remarkable and revolutionary advances in understanding our place in the Universe. Perhaps foremost amongst these was the transformative understanding of the motion of the Earth, and the associated acceptance of the heliocentric hypothesis. Another challenging task which underpinned the field for centuries, and which continues to do so to this day, was that of determining distances to the stars. When quantified for the first time in the 1830s stellar distances revealed, at a stroke, the utter vastness of the Universe. 

In the remainder of this introductory section, I will provide some background for those less familiar with astronomy. In particular, I will outline some of the broader issues associated with measuring angles, and the underlying scientific objectives which motivate their measurement. I will not dwell on how these observations are made in practice.

\subsection{The Measurement of Angles}

Astrometric measurements essentially involve determining the position of a star (or other celestial object) as it appears projected on the celestial sphere (the night sky). The star's distance being, at least to a first approximation, unknown, positions at any time can be simply and uniquely specified by the two angular coordinates of spherical geometry, precisely corresponding to latitude and longitude on Earth. The origin of the chosen coordinate system is a delicate issue in practice, but the principle is straightforward: just as for geographical latitude, one coordinate can be tied to the extension of the Earth's equatorial plane and, as for geographical longitude and the choice of the prime meridian as its origin, the other is referenced to some arbitrary but well-defined direction in space.

The basis of astrometric measurements, then, is the accurate measurement of tiny angles that divide up the sky. Dividing a circle, whether on paper or on an imaginary sweep of the celestial sky, is a task well-posed in principle.  Practical techniques for doing so aside, it is only necessary to agree on the unit of subdivision. Although scientific users today work and calculate angles in radians, the commonly accepted choice of three hundred and sixty degrees in a circle was made for us long ago. Ascribed to the Sumerians of ancient Babylonia, more than 2000~BCE, it was perhaps guided by the number of days in a year. One degree was subdivided into sixty minutes of arc, and each minute of arc was divided still further into sixty seconds of arc. The choice of sixty rests on the number itself being highly composite: it has many divisors, which facilitated calculations with fractions performed by hand. 

To visualise these angles, the Sun and the Moon both cover the same {\it angle\/} on the sky, about half a degree. The much smaller one second of arc corresponds to a linear distance of one meter viewed from a distance of about two hundred kilometers. This very small angle turns out to be a particularly convenient angular measure and benchmark in astronomy, and it has been used to construct the very basic measure of astronomical distances, the parsec (described further below). In very round numbers, one second of arc is also the angle to which astronomers can measure, with relative ease today, the position of a star at any one moment from telescopes sited below the Earth's atmosphere. It is the shimmering atmosphere which has most recently pushed these measurements to be made from space, and a little background is useful to explain more carefully why.

The lowest portion of our atmosphere is known as the troposphere (from the Greek `tropos' for `turning' or `mixing'). Extending to a height of about ten~kilometers, it contains three quarters of the atmosphere's mass, and within it turbulent mixing of the air, due to convective heating rising from the Earth's surface, plays an important part in our atmosphere's structure and behaviour. Turbulence affects light rays passing through the atmosphere, and causes the familiar twinkling of star light. Already Eratosthenes (276--194~BCE) had commented on their `tremulous motion' \citep{1970fraser}. To minimise these effects, astronomers build their telescopes at high mountain sites where the thinner atmosphere and smaller turbulence effects gives more stable images (as well as being away from the bright lights of roads and cities). At good sites, the dancing motion might drop below a second of arc, but not by much more. 

The human eye imposes its own limit to measuring angles of about one minute of arc. Mainly determined by the small diameter of the pupil through which light enters the eye, this limit is many times worse than that imposed by the atmosphere. Until the invention of the telescope, observations by eye had therefore placed considerably less stringent limits on the accuracy of star positions. The introduction of the telescope, credited to Dutch opticians in the opening years of the seventeenth century, but considerably and more famously improved upon by Galileo in 1609 \citep[e.g.][]{1955king,1956rosen,1976ilardi,1977JHA.....8...26V,1977TrAPS..67.....V}, brought with it two distinct improvements. First, it extended the faintness limit of the stars that could be seen, revealing countless more than were visible by eye. The larger diameter of the telescope aperture also gave an improved accuracy of positional measures. 

Making telescope mirrors larger improves the accuracy proportionately, but only up to the point that the atmospheric turbulent motion sets in at around one second of arc \citep[e.g.][]{1966JOSA...56.1372F}. Larger than that, even very large telescopes on the ground, the largest now reaching a diameter of ten~meters, with even larger under planning, fail to break through the accuracy limit on angular positions set by the atmosphere. For this specific reason, there is an important distinction between the angular resolution of an optical telescope (given approximately by $1.22\,\lambda/D$, where $\lambda$ is the wavelength and $D$ the mirror aperture) and the resulting formal accuracy in positional location \citep{1978moas.coll..197L} on the one hand, and the relative positional accuracies that can be achieved over large angles across the sky on the other \citep{1980A&A....89...41L}. We may also note in passing that significantly better {\it relative\/} positional accuracies can be achieved across very small angular fields of view (see also Section~\ref{sec:narrow-field}), while at radio wavelengths aperture synthesis making use of the Earth's rotation can also achieve high relative positional accuracies on a very small number of intense radio-emitting stars \citep[e.g.][]{1999A&A...344.1014L}. 

So much for one second of arc. It is a tiny angle, corresponding to the size of a Euro coin viewed from a distance of 5\,km. It remains problematic enough to measure, and one which proved a great challenge for the astronomical instrument makers of earlier centuries. But space astrometry has now reached accuracies of one thousandth of one second of arc. It is an angle that corresponds to the size of an astronaut on the Moon viewed from Earth, a golf ball in New York viewed from Europe, the diameter of human hair seen from 10\,km, or the (angular) growth rate of human hair in one second when viewed from a distance of 1\,m. The next advance in space astrometry, specifically the Gaia satellite due for launch in 2013, plans a further advance of a factor one hundred, targeting accuracies of a few microseconds of arc, corresponding to one Bohr radius viewed from a distance of~1\,m.  Such accuracies naturally pose extreme engineering challenges for optical quality, detector performance, and gravitational and thermal instrumental flexure.

\subsection{The Relevance of Star Positions}

Measuring an accurate position {\it per se\/} is rarely the ultimate objective of astrometry. Rather, the positions of stars in the sky vary minutely with time for a number of reasons. The crucial point is that repeatedly measuring a star's position over a period of months and years can track certain tiny motions which prove central to understanding their nature. It turns out that measuring the positions of stars offers deep insight into their properties, with cascading implications on diverse topics such as the structure and origin of our Galaxy, and the origin and age of the Universe.

Those with even a little familiarity with the night sky will know that the stars do not appear in the same position from night-to-night, nor even between the start of the night and the end, but {\it appear\/} to move slowly across the night sky. To first order, the stars occupy fixed positions relative to each other, and it is simply the spinning of the Earth on its axis, one rotation every twenty four hours, combined with its motion around the Sun, once per year, which together give the stars their apparent collective movement.  The apparent systematic rotation of the heavens is a direct consequence of the rotating and moving Earth.

Every star is moving through space. We know this now, although three hundred years ago scientists did not. As a result, over many decades or centuries, small displacements of some of the most swiftly moving stars do begin to be discernible. The manifestation of these star motions was first reported by Edmond Halley in 1718. Following Halley's discovery, other stellar motions were soon reported, and the collective study of stellar motions was born. Today, ultra-high accuracy astrometric measurements allow these star motions through space to be detected and quantified over relatively short periods of time of months or years. Nevertheless, as far as the human eye is concerned, relative star positions remain fixed without change over hundreds if not thousands of years. 

Measuring how the angular position of each star on the celestial sphere changes with time gives what astronomers call the star's `proper motion'. The name is a little cryptic, probably drawn from the French `propre' for `own', but was used to make clear that what is being measured is the motion through space of the star itself, and not that due to other effects like the Earth's rotation. It also reminds us that what has been measured is an {\it angular\/} shift over time, and not the actual speed of the star through space. 

The two are closely related, but the distinction is important. What we see is simply the star's movement projected onto the celestial sphere, which we can only describe in terms of an angular motion. Without knowledge of the star's distance, its true velocity through space cannot be inferred: a star whose position has changed by a certain amount over a few years might be a relatively nearby star moving slowly through space, or a star at a greater distance moving more rapidly. In practice, stars with large proper motions do tend to be nearby, and searching for high proper motion objects, by comparing photographs of the sky taken a few years apart, has proved to be a bountiful way of sifting out potentially nearby stars. Most stars, meanwhile, are far enough away that their angular proper motions are very small. 

Knowledge of a star's distance is therefore needed to convert its angular motion projected on the celestial sphere into a true space velocity. And knowledge of a star's distance is needed to convert its observed properties, and in particular its apparent brightness (which varies considerably between stars), into true physical quantities, notably its intrinsic luminosity. It is these basic physical properties of each star which are essential ingredients in putting together a picture of its composition and its internal structure, its age and its past and future evolution.

\begin{figure}[t]
\centering
\includegraphics[width=0.7\linewidth]{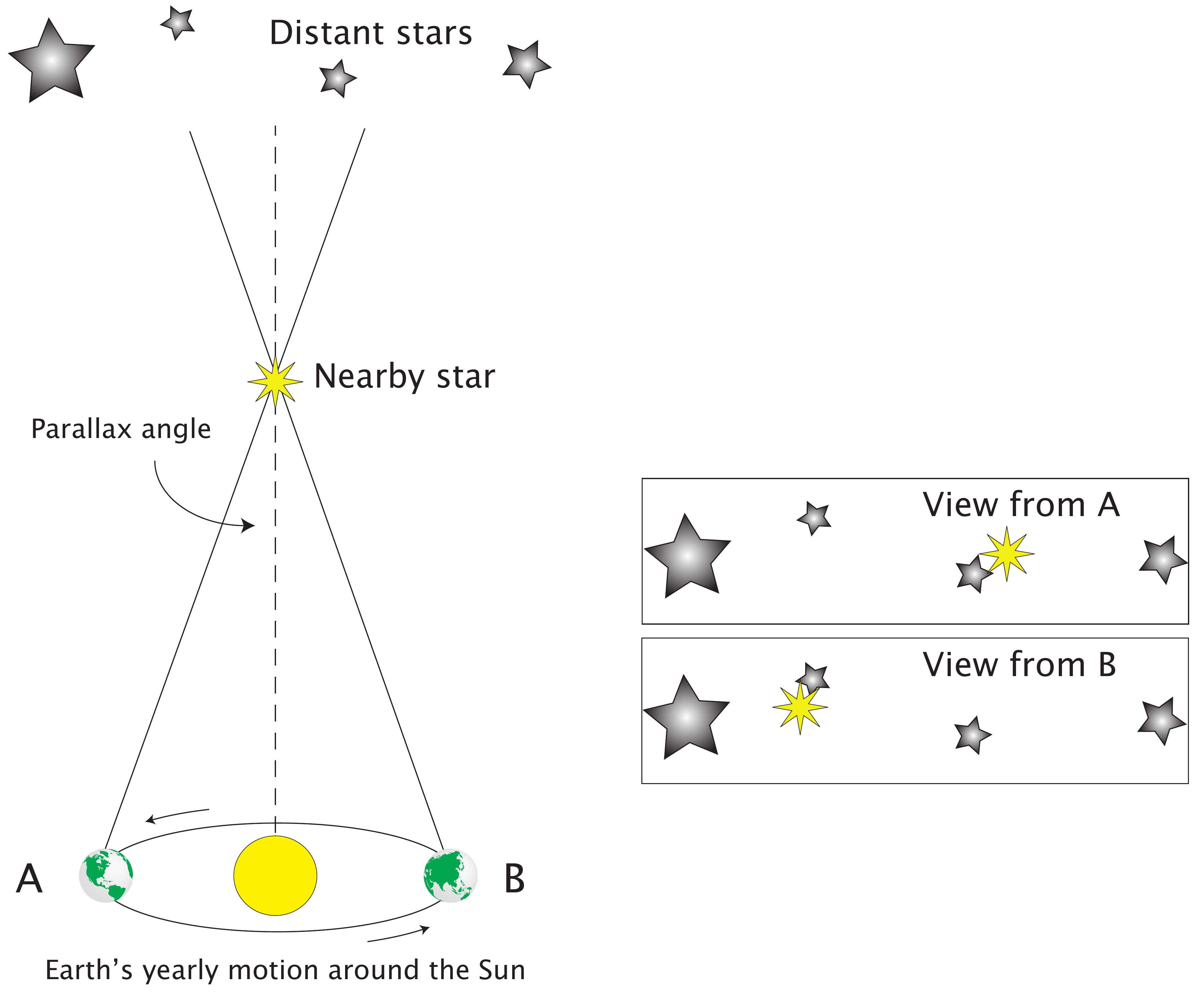}
\caption{\footnotesize Parallax measurement concept. As the Earth moves in its annual orbit around the Sun, nearby stars appear to be displaced differently with respect to the more distant background stars. The angles in this schematic are greatly exaggerated.
\label{fig:parallax}
}
\end{figure}

Here we encounter a major problem, for the distances to even the nearest stars are truly vast. They are enormously and extravagantly large, and there are no analogies that really allow us to comprehend them.  John Herschel (1792--1871), son of the illustrious William, attempted to describe the unimaginable distance scales \citep[quoted in][]{1899allen}: {\it ``To drop a pea at the end of every mile of a voyage on a limitless ocean to the nearest fixed star, would require a fleet of ten thousand ships, each of six hundred tons burthen.''} It's an inconceivable distance, and an equally inconceivably meaningless number of peas.  In another attempted analogy, the world's population of seven billion people, spaced out one every five thousand kilometers, would just about stretch to the nearest star. This preposterous extent of space, but more poignantly its stark emptiness, is perhaps better conveyed by a scale model in which the Sun is shrunk to a marble one centimeter in size: the Earth would be a grain of salt one meter from it, and Pluto would sit far beyond at forty meters. In this grand orrery, the {\it nearest\/} stars, Proxima and Alpha Centauri, would be two hundred kilometers away.

The key to measuring stellar distances is actually based on the classical surveying technique of triangulation. It is based on the fact, known since the time of Copernicus, that the Earth moves around the Sun, taking one year to complete its orbit.  This yearly motion provides slightly different views of space as we move around the Sun. The nearest stars then {\it appear\/} to move back and forth with respect to the more distant ones over this annual cycle (Figure~\ref{fig:parallax}). The problem is that the back-and-forth motion is minuscule. The picture of the grain of salt orbiting a marble at a distance of a meter, and using this perspective change to measure a point of light two hundred kilometers away, describes the challenge. In practice, astronomers struggled for centuries to measure the first stellar distances, not so surprising in view of the colossal (and unknown) problem facing them.

In terms of the Earth's orbital motion around the Sun, then, each star has its own `parallax angle', corresponding to the ratio of the Earth--Sun distance to that of the star. If measured during this yearly motion, nearby stars appear to oscillate slightly more, back and forth, compared to the more distance stars. The underlying principle of measuring stellar distances, then, is actually rather straightforward; it's just the small size of this parallax motion that makes the task so challenging.

We now know that the stars nearest to the Sun, for example Alpha and Proxima Centauri, have parallax shifts of around one second of arc, while more distant stars have smaller parallax angles still.  Down the centuries, attempts to measure this parallax effect failed repeatedly because the relevant angles were so small, almost as if the stars were points of light at infinite distance. The effect was first measured, finally, only in the 1830s.

It's another of Nature's coincidences that atmospheric blurring is about the same size as the parallax shift of nearby stars. Measuring distances to more distant stars beyond our immediate neighbours required accessing parallax angles of one tenth or one hundredth of a second of arc, and this became feasible only during that latter part of the twentieth century. But measuring more distant, larger numbers, or rarer types of stars, requires measurement accuracies of around one thousandth of a second of arc. Before the advent of space measurements, such goals remained firmly beyond reach.

The parallax-based distance measurement technique is so basic that the fundamental unit of distance measurement in astronomy beyond the solar system is based upon it. Conveying the essentials of `parallax' and `second' of arc it is referred to as the {\it parsec\/} (the name was proposed in 1913 by H.\,H.~Turner, and was subsequently adopted by Commission~3 of the International Astronomical Union in 1922; some of the associated debate is recounted by \citet{1913Obs....36..160.}). One parsec is simply the distance at which a star has a parallax angle of one second of arc as the Earth moves in its annual orbit around the Sun. The light-year is another convenient description of distance in astronomy, and the two units are often used side-by-side (although only the parallax can be measured directly).  The light-year is the distance covered by light, which travels at nearly three hundred thousand kilometers a second, over a time interval of one year. Putting in the numbers, one parsec is a little more than three light years, and one light-year is some ten million million kilometers. The nearest stars to the Sun are at a distance of a little more than one parsec, or around four light-years. In terms of parallax angles then, astronomers needed to master measurement accuracies of around one second of arc to measure the distances to the nearest stars. And the more distant the star, the smaller the parallax. 

The nearest (gravitationally bound) star cluster to our Sun, the Hyades, lies at a distance of around forty parsecs, or a little more than a hundred light years. The spiral arms of our Galaxy closest to us are at around five hundred parsecs, and the centre of our Galaxy is nearly ten thousand parsecs distant, or a colossal thirty thousand light years.  Our nearest neighbouring galaxies, the Magellanic Clouds, are some fifty thousand parsecs. Beyond that, great galaxy clusters stretch out to distances of tens of millions of parsecs or more---their light taking tens of millions of years to reach us.

\begin{figure}[t]
\centering
\includegraphics[width=0.55\linewidth]{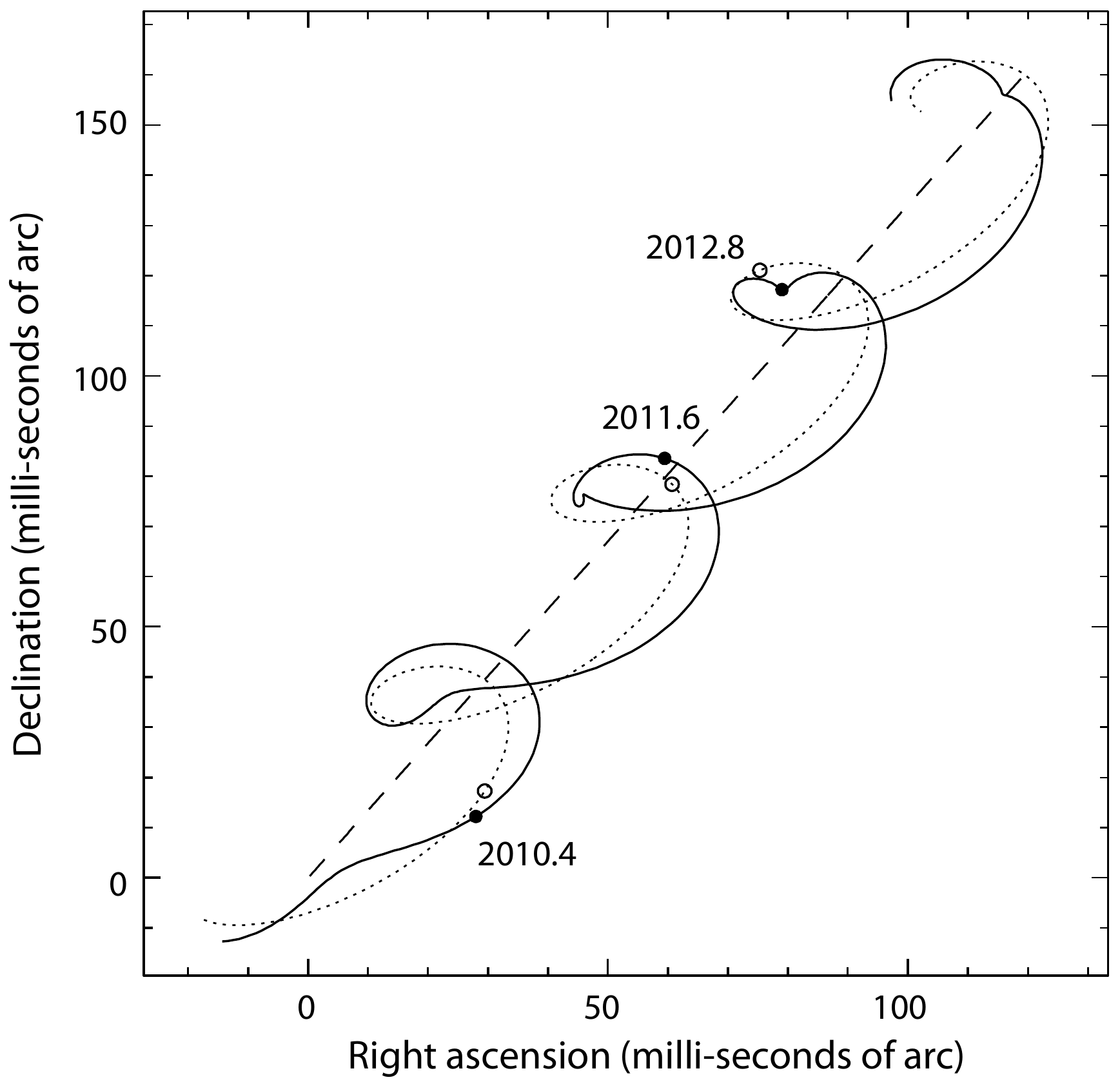}
\caption{\footnotesize The figure illustrates the principle of how the (angular) position of a star on the celestial sphere would be perceived to change with time, as a result of various physical effects, when viewed at extremely high angular resolution. Far from being fixed in one direction, a typical star moves through space at a velocity of some 10--30~km per second, tracing out a linear path across the sky with time (dashed line). Superimposed on this, the star's position {\it appears\/} to oscillate due to the Earth's motion around the Sun, providing access to the star's distance through the effect of parallax (dotted line, which shows this `reflex' motion over roughly four years). A planet in orbit around the star results in a yet more complex motion of the star around the star--planet system's centre of mass (solid line).
\label{fig:astrometric-path}
}
\end{figure}

We can summarise our present knowledge, and therefore the context in which astrometric measurements have advanced through history. If viewed under a celestial magnifying glass, each star in the sky has a tiny component of angular motion due to its velocity through space, plus a minuscule apparent oscillatory motion due the Earth's annual motion around the Sun, along with various other higher-order components of motion. These include the effects due to binary companions (which can exceed the effect of parallax), planets in orbit around them, and general relativistic light-bending. Quantifying these are the goals of contemporary astrometry (Figure~\ref{fig:astrometric-path}).

\section{Early History}

\subsection{Developments in Ancient Greece}

The flourishing of western astronomy over the past few hundred years has its origins in much earlier bursts of scientific activity. Prehistoric sites revealing celestial alignments, such as Newgrange in Ireland and Stonehenge in England, date from around 3000~BCE \citep{1997ruggles}. 

The first recorded developments emerged in Mesopotamia around 1000~BCE where, in the land between the Tigris and Euphrates rivers now occupied by southern Iraq, Assyro--Babylonian astronomers systematically observed the night skies, building on common lore already conscious of the changing daylight over the year. They observed, measured, and recognised, for the first time, that certain celestial phenomena were periodic:  amongst them the regular appearance of Venus, and the eighteen year cycle of lunar eclipses. Their careful records formed the basis for later developments, not only in ancient Greek and Hellenistic astronomy, but also in classical Indian and medieval Islamic astronomy. 

Early Greek philosophers, the Pythagoreans amongst them, played a key part in astronomy's earliest awakening \citep{1932heath,1997hoskin}. They believed that the underlying regularities, or laws of nature, were discoverable by reason. As part of this philosophical school, astronomers of ancient Greece tried to understand the Universe based on principles of `cosmos', or order. The revolutionary idea that the Earth might be spherical began to replace the pre-Socratic view that its surface was flat \citep{1990crowe}. 

Plato (427--347~BCE) and his contemporaries knew that the heavens rotated night after night with constant speed, the `fixed' stars preserving their relative positions as the heavens turned. Moving amongst them in a complex and unfathomable way were the seven wanderers---the Greek {\it planetes}---the Sun, the Moon, and the planets visible to the naked eye: Mercury, Venus, Mars, Jupiter and Saturn.  Seen from Earth, their positions trace out complex and convoluted paths, sometimes with even backward `retrograde' loops. As described by \citet{1991goodman-russell}: {\it ``Their erratic behaviour had baffled and infuriated generations of Greek thinkers, up to Plato himself. It seemed impossible to reconcile their celestial meanderings with either the supposed divinity of heavenly bodies or with any simple concept of circular motion.''}

Scientific thinking was dominated by the idea that the Earth lay fixed at the centre of the Universe. This fundamental tenet in mankind's early views completely obstructed the correct interpretation of planetary motions. We now know that the apparently complex paths of the planets follows from the rotation of the Earth, combined with the orbits of the Earth and other planets around the Sun. When interpreted correctly in a heliocentric system, and with elliptical rather than circular orbits, the motions are simple. But in a system in which the Earth is fixed they are not. Heraclides had hinted at a Sun-centred system in the fourth century~BCE, but his view failed to find support in a culture generally attached to the idea of an Earth fixed in space, which would continue to hold sway, erroneously, for a further two millennia. Perhaps we should really not be too surprised by the reluctance of the ancient Greeks to accept our present (rather non-intuitive) idea of a rapidly spinning Earth careering through space.

Aristarchus of Samos (circa 310--230~BCE) made one of the first attempts to determine the distances and sizes of the Sun and Moon \citep{1913heath}. He deduced the ratio of their distances using trigonometry, by measuring the angle between them when the Moon is exactly half lit. He also argued in favour of the heliocentric, Sun-centred system, a view supported by Seleucus of Seleucia around the second century BCE. But these ideas found little favour at the time, and they remained lost amongst the geocentric, Earth-centred system still being championed by most of his contemporaries. 

To explain the complex apparent motions of the planets and the varying speed of the Moon, geocentric proponents could not appeal to planetary orbits which were simply circular. They had to introduce complex epicyclic motions---patterns traced out by circles turning around the circumference of larger circles (Figure~\ref{fig:macp-epicycles-and-ellipses}a). Contrived though they were, they broadly explained the irregular speeds of the planets across the sky throughout the year, occasionally even tracing backward loops with respect to the background stars. 

At around the same time, Eratosthenes (276~BCE--194~BCE) invented a system of latitude and longitude, and used the varying elevation of the Sun to estimate the size of the Earth, deriving a value which would be used for centuries afterwards. He argued as follows: on the summer solstice at local noon in Swenet (modern day Aswan) the sun appeared at the zenith, while in his home town of Alexandria, assumed to lie due south, the angle of elevation of the sun was 1/50th of a great circle south of the zenith at the same time. He concluded that the distance from Alexandria to Swenet must therefore be 1/50 of the total circumference of the Earth.

Hipparchus (active circa 160--126~BCE) is credited with a number of important advances in astronomy, although most of what is known about his work is handed down from Ptolemy's second century thirteen-volume {\it Megale Syntaxis}, or `Great Compilation'. This became better known subsequently as the {\it Almagest}, `The Greatest', under the name assigned by ninth century Arabic translators \citep{1984toomer}. Ptolemy pioneered the classification of star brightnesses still in use today, dividing them into six groups, the brightest designated as first magnitude (the first to be seen at dusk), and the faintest as sixth. He followed the ancient Babylonians in dividing a circle into three hundred and sixty degrees, each of sixty minutes of arc, and he compiled the first systematic star catalogue, recording star positions with an accuracy of about one degree. He was the first to describe the precessional motion of the fixed stars, that is the steady wobbling of their positions over decades due to the steady change in the position of the Earth's spin axis in space, just like a spinning top.

But Hipparchus incorrectly continued to uphold the geocentric system. His argument was that a precisely circular orbit of the Earth around the Sun failed to explain the planetary motions. We know now that the planetary orbits are elliptical, so that his argument was compelling, but fallacious. Nevertheless his views, and his considerable authority, effectively ensured that the heliocentric hypothesis would lay discarded for many centuries. More than two hundred years later, in the second century~CE, Ptolemy would put forward his own variant of this geocentric world view, and would also invoke the complex epicyclic motions to predict, successfully even if based on flawed models, the future positions of the planets.

Greek scientific activity came to a reasonably abrupt end. According to \citet{1991goodman-russell}: {\it ``\ldots the most likely reasons seem to be the paucity of scientists and their isolation. The Athenian institutions and the Alexandrian Museum were rarities. Over the Greek centuries most scientific activity was uncoordinated, and poor communications often resulted in duplication of effort or ignorance of what had been achieved. Education in Greek schools concentrated on music, poetry and gymnastics, not on science; with the exception of Alexandria there was no strong government or social encouragement of the sciences. For Europe to have developed the sciences further from these Greek foundations, knowledge of Greek, close contact with the Greek scientific texts, and sustained interest in what they might teach were all necessary. But in the centuries after the fall of the Roman Empire in the west, none of these conditions was satisfied.''}

\subsection{The `Dark Ages' in Europe}

The subsequent decline of the Roman empire---in population, economic and political order---precipitated by barbarian attacks, decimating epidemics, and inability to provide for the succession of government---ushered in the `Dark Ages'. Europe disappeared under the likes of the Vandals and Visigoths for centuries.

To the East, meanwhile, China's first major economic burst under the Han dynasty (206~BCE--220~CE) nurtured a philosophical period roughly coincident with the innovative centuries of Greek philosophical and scientific thought. 
In India the Gupta empire, from around 320~CE, also stimulated navigation and advances in numeracy, embracing the concept of `zero' as well as the use of Arabic numbers. Astronomy was recognised as a separate discipline, and around 500~CE Aryabhata held that the Earth was a sphere rotating on its axis. Much later in China, under the Sung emperors in the eleventh and twelfth centuries, an advance in scientific achievement was accompanied by the flourishing of observational astronomy. Patronage from the emperors ensured the astrological fortunes of their dynasty, as well as regulation of the official calendar.  Star catalogues, as well as records of sunspots and comets, have been handed down to us from this time. But China's separation from the west, buffered by the nomadic tribes of central Asia, meant that their developments (such as their records of planetary motions, novae, and supernovae) had little immediate influence on Europe's subsequent scientific re-awakening.

The burgeoning Islamic culture was to dominate the world's economic development from the early part of the seventh century for the next three hundred years \citep{1994saliba}. Geographically closer than China and India, it had more of a direct influence on the west, and played an important part in reviving scientific enquiry in Europe. Supported by the patronage of the Caliphs, Islamic scholars transmitted, translated, and criticised anew the ancient Greek scientific texts. Knowledge of astronomy was also inspired by practical needs: to establish each mosque's direction to Mecca, the timing of daily prayers, and to rule on the precise beginning and end of Ramadan. 

\begin{figure}[t]
\centering
\includegraphics[width=0.55\linewidth]{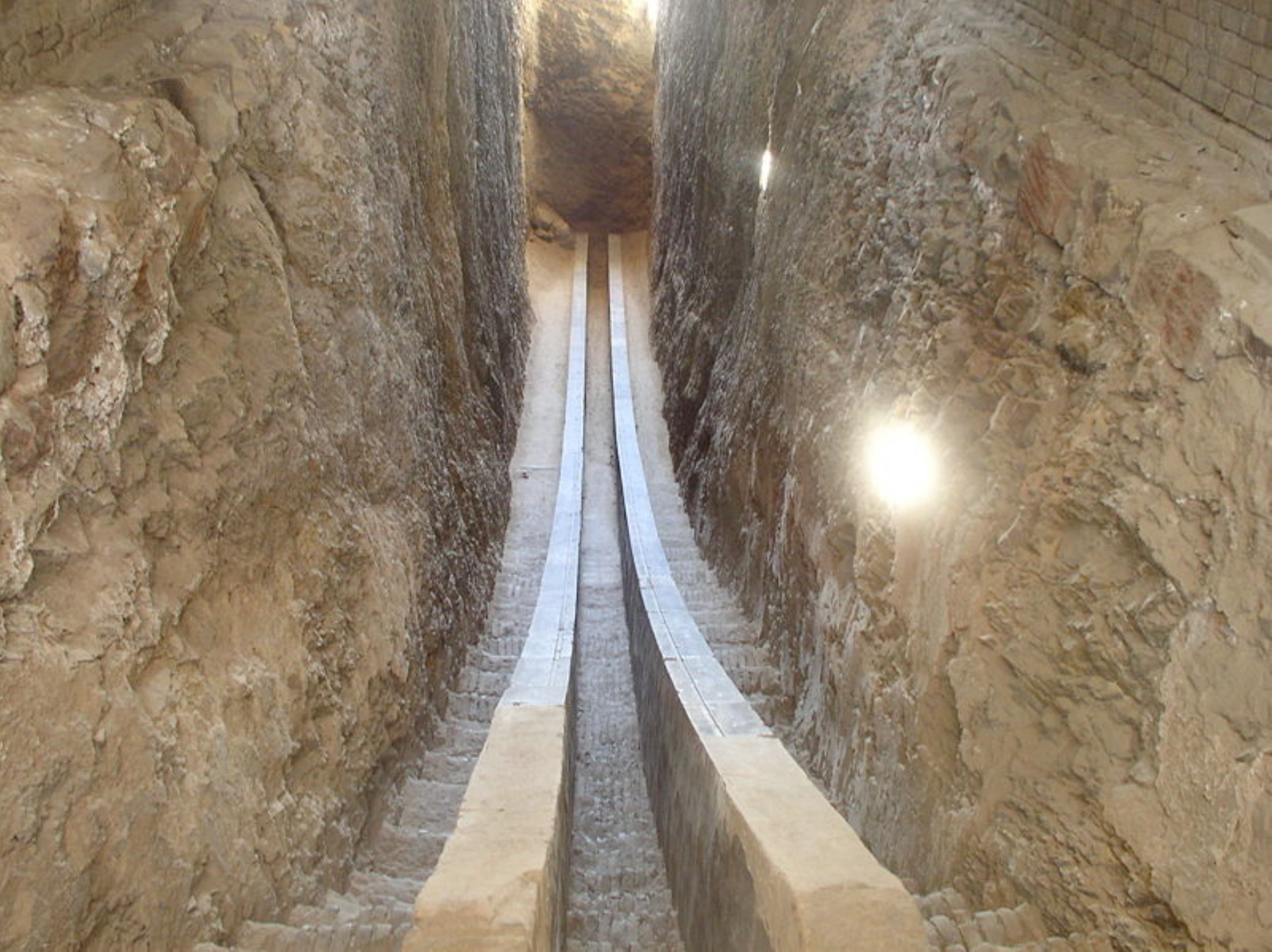}
\caption{\footnotesize Ulugh Beg's observatory, Samarkand. His 36-meter radius sextant, dating from 1428, was sited in a circular arc between these marble walls (courtesy Alex Ostrovski).
\label{fig:ulugh-beg-observatory}
}
\end{figure}

Amongst notable achievements in astronomy, Al~Battani, working around 900~CE at his observatory on the Euphrates, refined Ptolemy's description of the orbits of the Sun and Moon. Ibn Yunus (circa 950--1009) described planetary alignments and lunar eclipses accurate enough for a great figure in late nineteenth century astrometry, Simon Newcomb, to use them for his own theories of lunar motion. Ulugh Beg, grandson of the Mongol conqueror Tamerlane, constructed a sextant of thirty-six meter radius in Samarkand in 1428, sited in a circular arc between marble walls (Figure~\ref{fig:ulugh-beg-observatory}). He used it to compile a catalogue of 994 stars.  With star positions  accurate to about one degree, it is generally considered as the greatest star catalogue between those of Hipparchus and Tycho Brahe. Further details of these early astronomers and their achievements, amongst others, can be found in \citet{2007BEofA}.

\subsection{European Revival}

Europe's slow emergence from the `Dark Ages' began to gather pace under the Carolingian dynasty, from around 730~CE onwards. Trade and towns in western Europe started to revive, and economic life progressively shifted from the Mediterranean to the North Sea and Atlantic coast. From this renewed prosperity, and improved political stability, the foundations of the modern age slowly emerged. Indeed a significant body of scholarship over the past fifty years has quite dispelled the idea that Europe between 500--1500~CE was intellectually and technologically stagnant. 

A new curiosity about the heavens surfaced and thrived. The basic imponderable of astronomy and cosmology until the Middle Ages, that of the inexplicable motion of the five planets known at the time, was picked up again after a pause of a full millennium. Nicholas Copernicus (1473--1543) openly drew on this rich medieval tradition, and finally laid the secure foundations for a credible heliocentric world model, in which the Earth moves in orbit around the Sun rather than vice versa. Little new observational evidence motivated his thinking: he lived before the invention of the telescope, and his best observational accuracy was only about ten minutes of arc. Rather, rediscovery and reinterpretation of the ancient texts played a major part in the origins of Renaissance culture in general, and astronomy in particular. According to \citet{1913heath}: {\it ``Copernicus himself admitted that the [heliocentric] theory was attributed to Aristarchus, though this does not seem to be generally known.''}

Copernicus proposed that the Earth, far from being fixed in space, was actually subject to three kinds of motion. The first was an annual orbit around the Sun. The second was a daily rotation accounting for day and night, but about an axis tilted with respect to its orbit plane which would account for the changing character of the seasons. Third was a more complex and very long period wobble of the Earth's axis as it spins, known as precession. His {\it De Revolutionibus Orbium Coelestium}, `Concerning the Revolutions of the Heavenly Bodies' \citep{1543copernicus}, arguably marks the beginning of Europe's scientific awakening. 

\begin{figure}[t]
\centering
\includegraphics[width=0.90\linewidth]{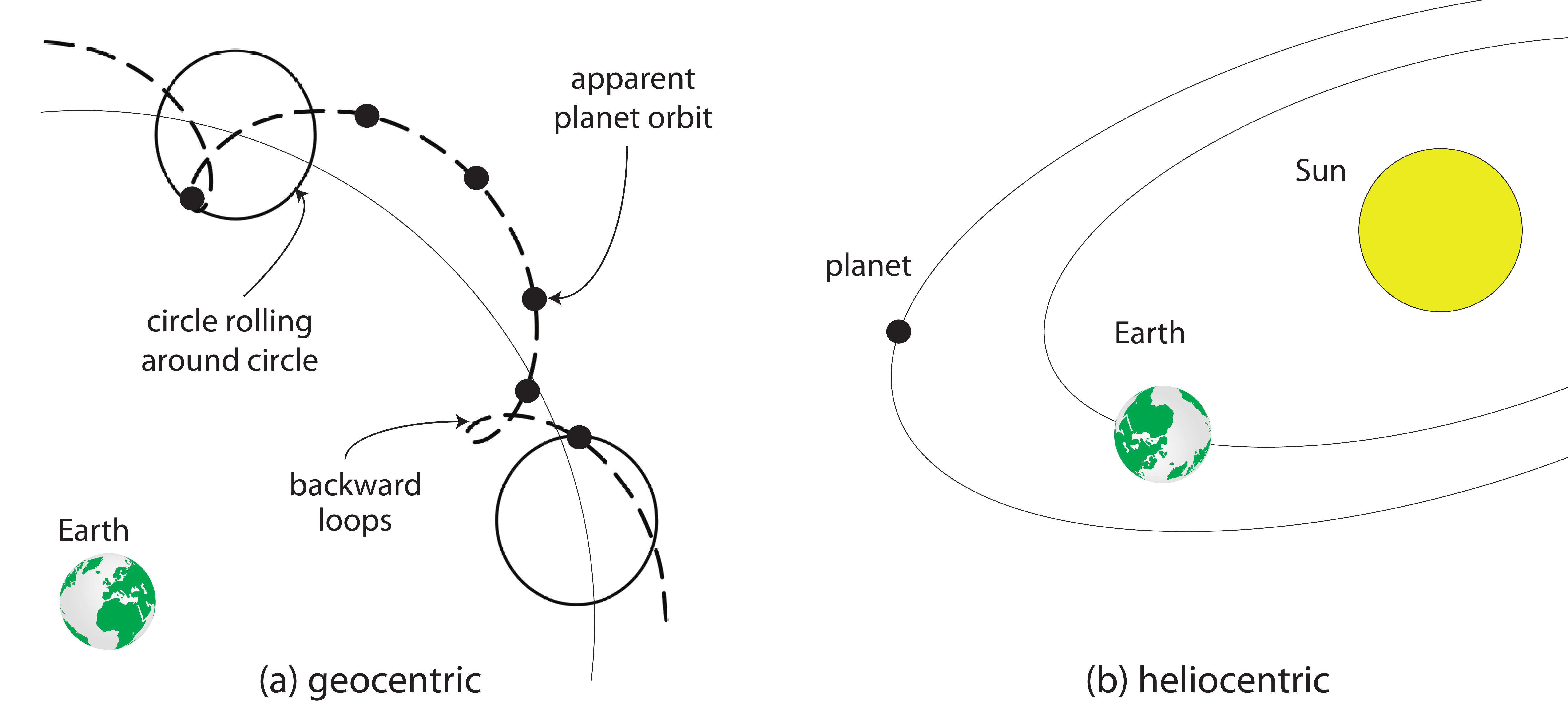}
\caption{\footnotesize (a)~In a (hypothetical) Earth-centred (geocentric) system, the apparent motions of the planets, as viewed from the Earth, could only be explained as a superposition of complex `epicyclic' curves. (b)~In our present understanding the planets, including the Earth, orbit the Sun in elliptical paths, and the solar system can be explained as a set of planetary masses orbiting the Sun according to Kepler's laws.
\label{fig:macp-epicycles-and-ellipses}
}
\end{figure}

The acceptance of a Sun-centred solar system accounted for the most extreme contributions to the backward looping motions of the outer planets. But Copernicus still needed highly contrived epicycles to explain the detailed planetary motions, albeit of a smaller magnitude than those invoked by Ptolemy in his Earth-based system (Figure~\ref{fig:macp-epicycles-and-ellipses}a). Yet other subtleties were needed to match the known orbits of the planets, for Copernicus was erroneously trying to fit a series of circular motions to their yet-to-be discovered elliptical paths. 

Only with the later work of Johannes Kepler, Galileo Galilei and Isaac Newton, and the realisation and understanding that the planetary orbits were elliptical rather than circular, could the need for epicyclic motions be discarded altogether (Figure~\ref{fig:macp-epicycles-and-ellipses}b). By demonstrating that the motions of celestial objects could be explained without putting the Earth at rest in the centre of the Universe, the work of Copernicus stimulated further scientific investigations and became a landmark in the history of modern science. 

In 1610, Galileo Galilei published his {\it Sidereus Nuncius} \citep{1610galileo}, which described the surprising observations made with his newly-invented telescope---mountains on the Moon, moons around Jupiter, and patchy nebulae for the first time resolved into innumerable faint stars. His support for Copernican heliocentrism set in train a lengthy and well-documented conflict with the Catholic Church, leading to suggestions of heresy, and his eventual trial and house arrest in 1633 \citep{1957drake}.  Giordano Bruno (1548--1600) was a proponent of heliocentrism and the infinity of the universe, earlier burned at the stake albeit for other more extreme theological heresies. Such were the harsh penalties for questioning the authority of the Holy Scriptures, which decreed that the Earth was the centre of the Universe, and that all heavenly bodies revolved around it. 

Johannes Kepler (1571--1630) appeared as an important figure in the seventeenth century astronomical revolution, best known for his eponymous laws of planetary motion. He defended heliocentrism from both a theoretical and theological perspective. His observational work with Tycho Brahe encouraged his own repeated attempts to calculate the full orbit of Mars around the Sun. Eventually, in 1605, he found that while a circular orbit did not match the observations, an elliptical one did. It was a simple answer which he had previously assumed to be too straightforward for earlier astronomers to have overlooked. He concluded that all planets move in ellipses, with the Sun at one focus. This deduction, his first law of planetary motion, provided a foundation for Isaac Newton's theory of gravitation. 

Isaac Newton (1642--1727) occupies a lofty pedestal unrivalled in the history of science, and his {\it Philosophiae Naturalis Principia Mathematica} of 1687 is arguably its most influential book.  He bestowed on mathematics and physics a rich and complex collection of ideas. Together, and in a stroke, his laws of motion, gravitational attraction, and the inverse square law of gravity gave an explanation of the motions of all celestial bodies. But this package of new ideas, Newtonianism, was not the only scientific movement competing for support, and it was not accepted immediately. Rivals included Hutchinsonianism in England, centred around the Trinitarian theology of John Hutchinson, and Cartesianism in France, based on the influential philosophical doctrine of Ren\'e Descartes.

The heliocentric hypothesis eventually prevailed, and Newtonian gravity along with it. With their joint acceptance came an inevitable consequence, a conclusion that would mark a fundamental turning point in science. For if the Earth indeed moves in orbit around the Sun, then the `fixed' stars cannot remain truly fixed in space. Unless they were at infinite distance, they would have to possess a parallax motion---an oscillation of their {\it apparent\/} position which would arise from the Earth's annual motion around the Sun (Figure~\ref{fig:parallax}). To be sure, neither Aristarchus nor Copernicus had observed the effect, and this fact alone implied that the distances to the stars must dwarf even the colossal distance scale of the solar system. 

The conclusion that the parallax effect had to exist therefore seemed inescapable. A renewed push to detect it began, armed with the certain knowledge that the effect being sought would be tiny. Great improvements in measurement accuracy would be needed before the effect could be measured. The warm-up was over, and the race to measure parallax began in earnest.

\subsection{Parallax, Newtonianism, and the Earth's Motion}

Starting some three or four centuries ago, the search for parallax, the further comprehension and definitive acceptance of Newtonianism, and understanding the precise nature of the Earth's motion through space were interwoven, and together motivated the progressive improvement of angular measurements. A related but more urgent practical problem came to a head at the same time: the navigational problems associated with the determination of longitude \citep[e.g.][]{1958water}.

For most of history, explorers in general and mariners in particular had struggled to determine their precise longitude, their point east or west of some reference point on the Earth. Latitude has the Earth's equator as a natural reference plane, and it can be determined by observing the altitude of the Sun or stars using specialised protractor-like instruments like the quadrant or sextant, or the astrolabe, a sort of analogue calculator capable of working out different kinds of problems in spherical astronomy \citep{1932gunther}. There is, however, no such unique reference position for longitude, and no practical means for its direct estimation. For a ship lost at sea on the slowly-spinning Earth, estimating longitude was frequently a matter of life or death. But it was tied directly to the knowledge of time. Without time, there was no hope of determining longitude: any uncertainty in the local time corresponds to an uncertainty in a star's transit across the local meridian, and an equivalent uncertainty in the observer's longitude.

The problem was urgent, and the economic consequence of ships, cargos and lives lost at sea was substantial. In France, Louis~XIV promoted the construction of the Paris Observatory, established in 1667 under director Giovanni Domenico Cassini, with the express purpose of extending France's maritime power and expanding her international trade. In England, King Charles~II was similarly moved to found the Royal Greenwich Observatory in 1675, with the purpose of compiling detailed star maps for navigational purposes \citep{1975forbes}. He instructed the first Astronomer Royal, John Flamsteed \citep{1997hoskin}, {\it ``to apply himself with the most exact care and diligence to the rectifying of the tables of the motions of the heavens, and the places of the fixed stars, so as to find out the so much-desired longitude of places for perfecting the art of navigation.''} In 1725 Flamsteed's {\it Historia Coelestis Britannica\/} was published posthumously, and contained his catalogue of 2935 stars \citep{1725flamsteed}. It was the first significant contribution of the Greenwich Observatory, and a landmark in the history of astrometry---positions, accurate to around ten or twenty seconds of arc, were the first measured with telescopic sights, and a major improvement over earlier work.

Star charts alone, however, could not provide a solution to the problem of navigation. Without a clock that could keep accurate time over months of an ocean voyage, there was no practical way of establishing what time it was at the reference point. With Galileo's discovery of the four brightest moons of Jupiter in 1610, named by him as the Medicean stars after his patron but subsequently named the Galilean moons in his honour, it became possible in theory to deduce the time on board ship by observing when the satellites appeared from behind the planet---the events occurred frequently and, more importantly, predictably. The world's first national almanac, the {\it Connaissance des Temps}, giving these eclipse timings, was published from 1679. Tables could then be consulted to see when these events were due to occur as measured at the prime meridian. Galileo himself pursued this approach to navigation during his lifetime, and even petitioned King Philip~III of Spain who had also offered a financial reward for a breakthrough in determining longitude. Yet such measurements could only be made at night, were much at the mercy of the weather, and quite impossible from a rolling boat in high seas, and it failed to provide a practical solution. Before the middle of the eighteenth century, most sailors continued to use a variant of dead reckoning to try to keep track of their position. Galileo died in 1642, before his method became widely used by cartographers on land. 

The search for a solution was spurred on by the Longitude Act of 1714, during the reign of Queen Anne. The British Parliament offered a prize of \pounds 20\,000, a fortune of some \pounds 6~million in present worth, for a method that could determine longitude within thirty nautical miles.  A solution was eventually found through the use of accurate celestial charts and lunar tables, in combination with the measurement of precise time. 

With the success of the marine chronometer in the 1760s, pioneered by English clockmaker John Harrison, time could at last be carefully measured and accurately transported throughout a long voyage. Accurate clocks eventually became commonplace. The problem of navigation at sea was considered as solved, and the Board of Longitude was dissolved in 1828 (the story is told in the popular account by Dava Sobel \citep{1996sobel}). Not until 1884, however, was the International Meridian Conference meeting in Washington~DC to adopt the meridian passing through Greenwich as the universal, if quite arbitrary and long contested, zero point of longitude. France abstained, maintaining her preferred use of the Paris meridian until 1911 for timekeeping purposes, and until 1914 for navigation. 

From 1767, the Nautical Almanac has been published annually. From 1958, the US~Naval Observatory and the HM Nautical Almanac Office have jointly published a unified volume, for use by the navies of both countries.  It still tabulates the positions of the Sun, Moon, planets, and a number of stars selected for ease of identification and widely spaced across the sky. To find the position of a ship or aircraft by celestial navigation follows the method unchanged for more than two centuries: the navigator uses a sextant to measure the height of a chosen star above the horizon, notes the time from a chronometer, and deduces location by comparing the star's position with that given in the almanac for that particular time. Thousands of lives and considerable fortunes had been lost before star charts in combination with transportable time could be used for reliable navigation. 

Well into the 1800s, star positions provided the most accurate means of determining geographical coordinates, and with them the distance between cities or the position of national borders. An interesting parallel occurs today: the huge civilian, commercial, and military reliance on global satellite navigation, notably GPS, depends crucially upon the inclusion of the delicate effects of Einstein's special and general relativity: omit them from consideration, and positions would be several kilometers in error after only a few hours. In this area alone, astronomy and relativity have proven indispensable to this important social and commercial venture.

As Copernicanism spread throughout Europe, and the heliocentric cosmos gained acceptance, the race to measure parallax gathered pace.  Even before 1600, astronomers were in agreement that the crucial evidence needed to detect the Earth's motion around the Sun was the measurement of trigonometric parallax. The early British Astronomers Royal, amongst others, appreciated the importance of measuring stellar distances, and had devoted much energy and ingenuity to the task. For example, according to \citet{1990chapman} {\it `Though the application of the telescope sight to angular measurement from the 1660s constituted a major technical breakthrough, the optics involved were simple, conservative in type, and secondary to the engraved divisions. This becomes apparent from the letters, notebooks, and Gresham College lectures of John Flamsteed, the first Astronomer Royal, for while his early decades at Greenwich were beset with instrumental problems, they were almost exclusively of a mechanical nature. The equality of scale degrees, or the regularity of a micrometer screw, claimed more attention than the resolving power of telescope lenses, and nowhere in his extensive writings is more than passing attention paid to optical resolution'.}

Indeed it was improved angular measurement, not enhanced visual acuity, that held the key to a range of astronomical problems from the sixteenth to the early nineteenth centuries \citep{1990chapman, 2001hirshfeld}. But it was still to take a further two hundred and fifty years, and failure upon failure, until the first star distances were measured. 

\subsection{The Advance of Astrometric Accuracy}

From ancient times through to the start of the twentieth century, the measuring of celestial positions had always been central to astronomical research. The quality of the instruments determined the accuracy of the measurements. The art of dividing a physical circular scale into degrees and minutes of arc was but a practical problem, essentially one of accurately marking off successively smaller angles. But it was one of such technical complexity that it now presented the principal barrier to advancing research.

\begin{figure}[t]
\centering
\includegraphics[width=0.8\linewidth]{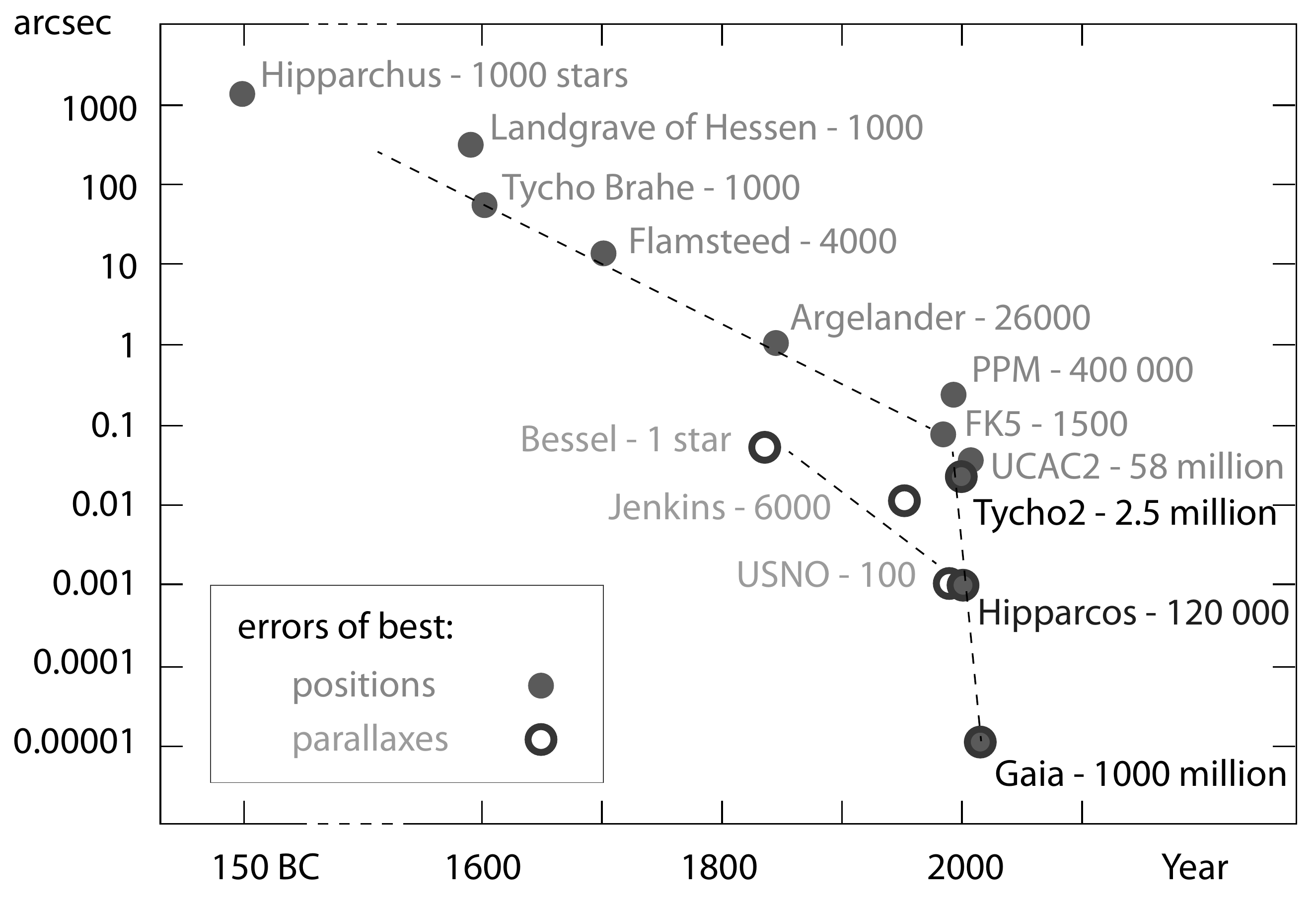}
\caption{\footnotesize Astrometric accuracy versus time (adapted from Erik H{\o}g). After several centuries of a more-or-less logarithmic improvement in accuracy with time, the advent of space-based measurement techniques (Hipparcos and Gaia) has led to an even more rapid improvement in accuracy. These space missions have also resulted in a unification in the accuracies achieved in star positions and parallaxes. 
\label{fig:accuracy}
}
\end{figure}

During the Middle Ages, European and Islamic astronomers adopted a brute force approach to the problem. They constructed observing circles with very large radii (and therefore very large physical dimensions) such that they could more easily inscribe and further dissect more precise angles on their annular limbs.  Tycho Brahe (1546--1601), whose observations provided the basis of Kepler's laws of planetary motion, employed such an instrument \citep{1598tycho,1921beckett,1990thoren}. His Great Quadrant had a radius of fourteen cubits, around seven meters, and probably reached an accuracy of around six minutes of arc, one fifth the Moon's diameter. At his lavish observatory of Uraniborg on the Danish island of Hven, developed under the patronage of Frederick~II, King of Denmark and Norway, he used his families of sextants, armillary spheres, and quadrants. By the last decade of the sixteenth century, he was reaching an unsurpassed accuracy of around twenty seconds of arc (Figure~\ref{fig:accuracy}). 

Despite his observational skills and his extravagant funding, Tycho attempted, but also failed, to detect parallax motion. But the accuracies that he achieved allowed him to deduce that the stars must lie several thousand times more distant than the Earth from the Sun. These distances were so immense that he was convinced Copernicus must be in error, and that the Earth was indeed fixed at the centre of a modified `Tychonic' system. In reality, with even the nearest stars having a parallax angle of only one second of arc, Tycho's accuracy was still twenty times too poor, and even his careful measurements could not but have failed to detect its effects. Nevertheless by the end of the sixteenth century, his catalogue of a thousand stars, and a similar effort by Landgrave (Baron) Wilhelm the Wise of Hesse (1532--1592) \citep{1618hesse}, set the standard for future surveys.

\begin{figure}[t]
\centering
\includegraphics[width=0.35\linewidth]{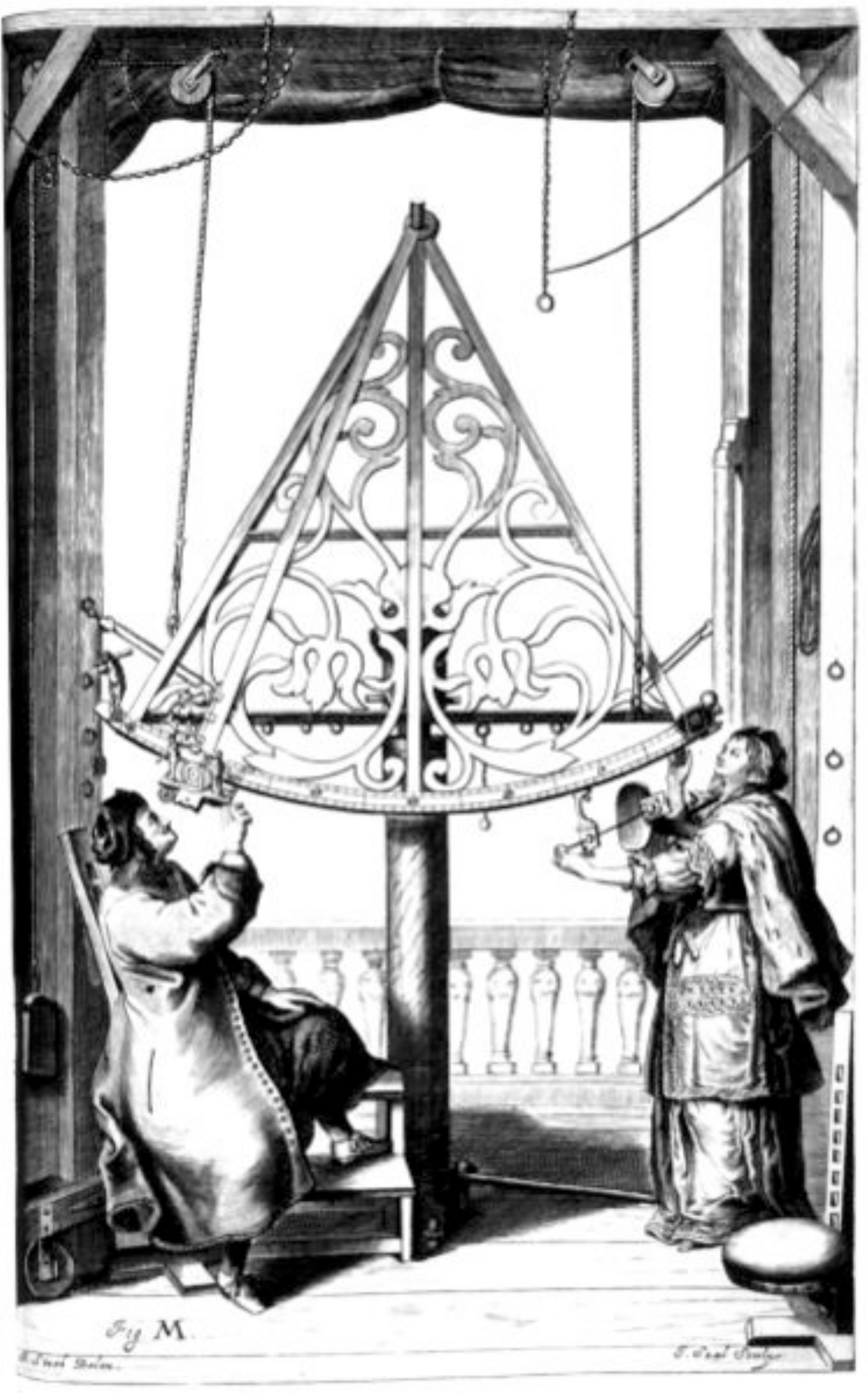}
\caption{\footnotesize The sextant of Johannes Hevelius \citep{1673hevelius,1971volkoff}; courtesy of Wikimedia Commons.
\label{fig:hevelius-sextant}
}
\end{figure}

The sextant and quadrant were protractor-like instruments designed to measure angles between pairs of stars, of up to sixty and ninety degrees respectively (Figure~\ref{fig:hevelius-sextant}). Catalogues were built up from many pairs of separations. Portable versions were later fixed in the meridian plane---the imaginary circle perpendicular to the celestial equator and horizon. Observations with wall-mounted `mural'  instruments began with Tycho's large meridian quadrant. Fixed to the local horizon, stars appear to drift past the local meridian as the Earth spins: this gave one part of the star's coordinates (the equivalent of geographical longitude, or right ascension) from the timing of its transit, and the other (the geographical latitude, or declination) from the graduated instrument itself. These were later replaced by the meridian circles, consisting of a horizontal axis in the east--west direction resting on fixed supports, about which a telescope mounted at right angles could revolve freely.

Until the late eighteenth century, the art of graduating circular scales into ever finer subdivisions was pursued in earnest, but carried out largely in secrecy to thwart the competition \citep{1955dewhirst,1972mckeon,1976chapman,1987bennett,1988chapman,1990chapman}.  Wider exposition of practical methods accelerated when the Board of Longitude, which had been formed in 1714 to solve the problem of finding longitude at sea, persuaded John Bird to publish his methods in 1767 \citep{1767bird}. In the following decades, Jesse Ramsden \citep{1777ramsden}, John Smeaton \citep{1786smeaton}, and Edward Troughton \citep{1809troughton} continued the advance of angular measurements. Prestigious Fellowships of the Royal Society were awarded for their instrument advances, underlining the importance with which the measurement of stellar positions was held, and testament to their innovation. In his chronicle of the rise and fall of economies throughout history, for example, \citet{2000jay} includes Ramsden's dividing machine for accurate graduation of circles for navigational and surveying instruments as one of the inventions which contributed to the productivity gain that signaled the Industrial Revolution.

During the later parts of the seventeenth and early eighteenth century, other instruments were added to the arsenal of techniques for measuring star positions. These included the transit telescope, which added a regulator clock to time the passage of stars across the Earth's meridian. Its more specialised form, the zenith sector, was used by Robert Hooke (1635--1703), one of the most important scientists of his age, in his own attempts to measure the parallax of the bright star Gamma Draconis \citep{1674hooke, 1950andrade,1956espinaase}.

Gamma Draconis is a giant star in the constellation of Draco, and a notable object throughout recorded history. According to \citet{1899allen}: {\it ``Its rising was visible about 3500~BCE through the central passages of the temples of Hathor at Denderah and of Mut at Thebes. And Lockyer [Sir Joseph Norman Lockyer, 1836--1920] says that thirteen centuries later it became the orientation point of the great Karnak temples of Rameses and Khons at Thebes, the passage in the former, through which the star was observed, being 1500 feet in length; and that at least seven different temples were oriented toward it. When precession had put an end to this use of these temples, others are thought to have been built with the same purpose in view; so that there are now found three different sets of structures close together, and so oriented that the dates of all, hitherto not certainly known, may be determinable by this knowledge of the purpose for which they were designed. Such being the case, Lockyer concludes that Hipparchus was not the discoverer of the precession of the equinoxes, as is generally supposed, but merely the publisher of that discovery made by the Egyptians.''}

The interest of the star Gamma Draconis to the seventeenth and eighteenth century parallax hunters was simply that it lay almost exactly in the zenith of Greenwich, minimising refraction by the atmosphere, and conveniently studied by a fixed telescope pointing straight up---Hooke had cut a hole in the roof of his apartment to observe it. In 1674 he claimed the detection of a parallax for Gamma Draconis of roughly thirty seconds of arc, and with it proof of the Copernican system, although later work showed that his results were in error. 

\subsection{Proper Motion and Stellar Aberration}

A remarkable and crucial breakthrough came in 1718. Edmond Halley, who had been comparing contemporary observations with those that the Greek Hipparchus and others had made, announced that the bright stars Aldebaran, Arcturus, and Sirius were displaced from their expected positions by large fractions of a degree \citep{1718halley}. He deduced that each star had its own distinct velocity across the line of sight, or proper motion. It was the first convincing experimental suggestion that stars were moving through space.

Halley's scientific achievements were many and varied \citep{1937macpike, 1998cook}. He predicted the return in 1758 of a periodic comet which now bears his name, identified solar heating as a cause of atmospheric turbulence, and suggested a measurement of the distance between the Earth and the Sun by timing the transit of Venus. Less successful was his suggestion, to explain anomalous compass readings, that the Earth was a hollow shell some eight hundred kilometers thick. This example also shows the limits in scientific understanding that existed a mere three hundred years ago. 

By 1725, instrumental advances had reached an accuracy of a few seconds of arc.  The Reverend James Bradley, England's third Astronomer Royal \citep{1963mccrea}, was deeply immersed in his own efforts to measure parallax, and was also focusing his attention on Gamma Draconis. His attempts were unsuccessful, for the star is too distant for the effect to show up at the accuracy then available. But they pushed his own estimates of the nearest stellar distances out to nearly half a million times that of the Earth from the Sun. 

More importantly, Bradley's experiments yielded an unexpected surprise: the detection of a small systematic shift in his star positions, of a form very different from that expected from the effects of parallax, and which he eventually correctly attributed as resulting from the addition of the velocity of light to the Earth's velocity as it moves in orbit around the Sun. The usual analogy is that when rain is falling straight down, and you're walking briskly ahead, you tilt an umbrella forward slightly to intercept the apparent direction of the rainfall.  It's a consequence of adding two velocities. Dinghy sailors know the effect well: the flag atop the mast doesn't indicate the wind direction, but that of the wind and boat speed combined. Bradley had pondered the meaning of his perplexing star measurements for three years before enlightenment struck, his insight precipitated by observing such a moving vane on a sail boat on the River Thames.

Bradley's observations of this effect, known as stellar aberration, or the aberration of starlight, was announced in 1729, and arguably rates as one of the most significant discoveries in the history of astronomy \citep[e.g.][]{1963blackwell,1963QJRAS...4...41M,1964stewart}. It provided the first direct proof that the Earth was moving through space \citep{1728bradley}. His results therefore supported the Copernican theory, that the Sun, rather than the Earth, was the centre of the solar system. But it confirmed, at the same time, Danish astronomer Ole R{\o}mer's discovery of the finite velocity of light fifty years earlier \citep{2005saito}. R{\o}mer had been observing the eclipses of Jupiter's moons as part of the ongoing challenge to establish a practical method to determine longitude \citep{1968VA.....10..105N}. His own conclusion that the velocity of light was finite, rather than propagating at infinite speed, wasn't fully accepted until Bradley's measurement of aberration provided crucial supporting evidence.

By failing to detect the parallax of Gamma Draconis, even at the unprecedented level of about one second of arc, Bradley's observations went further in confirming Newton's hypothesis of the enormity of stellar distances, and confirmed that the measurement of parallax would continue to pose a technical challenge of inordinate delicacy \citep{1966hoskin,1985vanhelden,1988Obs...108..199M}. In parallel with the direct search for parallax were less direct estimates of stellar distances, for example those made by Newton and others by appeal to the inverse square law, an approach resting on the simplistic (but incorrect) hypothesis that all stars had luminosities comparable to that of the Sun. This method was extended to circumvent the difficulties posed by the extremely bright Sun by the use of Jupiter as a (reflecting) intermediate calibrator, as first used for Sirius by James Gregory in 1668 \citep{1668gregory}, and for Vega by John Michell in 1767 \citep{1767RSPT...57..234M}.

Nevil Maskelyne, England's fifth Astronomer Royal \citep{1989howse}, spent seven months on the remote island of Saint Helena in 1761, a crucial staging and rendezvous point for sailing ships in the South Atlantic. He had been despatched by the Royal Society to observe the transit of Venus, and thereby to improve knowledge of the Earth's distance from the Sun and the scale of the solar system. He used a zenith sector and plumb-line in an unsuccessful attempt to measure the parallax of Sirius during the same expedition \citep{1970ashbrook}. 

During the eighteenth century, after Halley's first detection of stellar motions, the movements of many more stars were being announced. In 1783 William Herschel found that he could partly explain these collective motions by assuming that, in addition to the Earth's motion around the Sun, the Sun itself was moving through space \citep{1783RSPT...73..247H}. With his sister Caroline, Herschel made numerous important advances \citep{1954sidgwick}: he discovered Uranus in 1781, two moons of Uranus and two of Saturn between 1787--89, and discovered infrared radiation. He observed and catalogued binary stars, detecting the first orbital motions and, in the process, the first proof that Newton's laws of gravitation applied outside the solar system. He was a prolific telescope maker, and also sought to detect a parallax shift from measurements repeated over the course of a year. Yet in this, even armed with his largest telescope, a primary mirror more than a meter in diameter and a colossal twelve meter focal length, he too failed. As he wrote in 1782 \citep{1782herschel}: {\it ``To find the distance of the fixed stars has been a problem which many eminent astronomers have attempted to solve; but about which, after all, we remain in a great measure still in the dark.''}

Meanwhile, another important step in expanding ever larger star surveys was the work of J\'er\^ome Lalande (1732 --1807) in France. His {\it Histoire C\'eleste Fran\c caise} of 1801 \citep{1801lalande}, gave the places of 50\,000 stars with an accuracy of around three seconds of arc. 

The symbolic if arbitrary figure of one second of arc was now within sight, and attempts to measure parallax intensified. But since the distances to even the nearest stars were still unknown, nobody could predict what angular accuracy would be needed for the effect to be detected. The topic was the focus of many learned papers published in the opening decades of the 1800s \citep{1915Obs....38..249D,1915Obs....38..292D,1949PA.....57..259H,1975JRASC..69..153F,1975JRASC..69..222F}. The failures of Tycho, Hooke, Flamsteed \citep{1979JHA....10..102W}, Bradley, Maskelyne, Herschel and many others, were followed by a renewed flurry of measurements and false claims: amongst them by Giuseppe Piazzi in Palermo, Giuseppe Calandrelli in Rome, Fran\c cois Arago in Paris (later Prime Minister of France), Baron Bernhard von Lindenau in Gotha, Johan Schr\"oter in Lilienthal, and John Brinkley in Dublin. In the words of \citet{2001hirshfeld}: {\it ``Each claimed victory in what astronomers increasingly perceived as a parallax race. But instead of glory, the recent parallax competitors gained only the suspicion, if not the contempt, of their colleagues.''}

\subsection{The First Parallaxes}

What was urgently needed were criteria for selecting stars likely to be close to the Sun, to avoid time wasted in trying to measure distant stars. In 1837, German-born Wilhelm Struve \citep{1988batten}, working at Dorpat in Russia (now Tartu in Estonia), gave three suggestions: the star should be bright; it should be moving with a large angular rate across the sky (although this {\it could\/} be a rapidly moving star at a large distance, it was more likely to be `nearby'); and if the star was one of a binary pair, the two components should be well separated as judged by the time taken to orbit each other. Struve drew up a list of stars satisfying these criteria. Our present-day knowledge confirms that astronomers were, at last, able to select some of the very nearest stars on which to focus their painstaking measurements.

After many unsuccessful attempts, the very first stellar parallaxes were measured and reported during a burst of activity in the 1830s, two hundred years after Isaac Newton had removed any final doubt that the Earth was in motion around the Sun. After this protracted marathon to detect the first parallax, three scientists breasted the winning tape almost together.

\begin{figure}[t]
\centering
\includegraphics[width=0.45\linewidth]{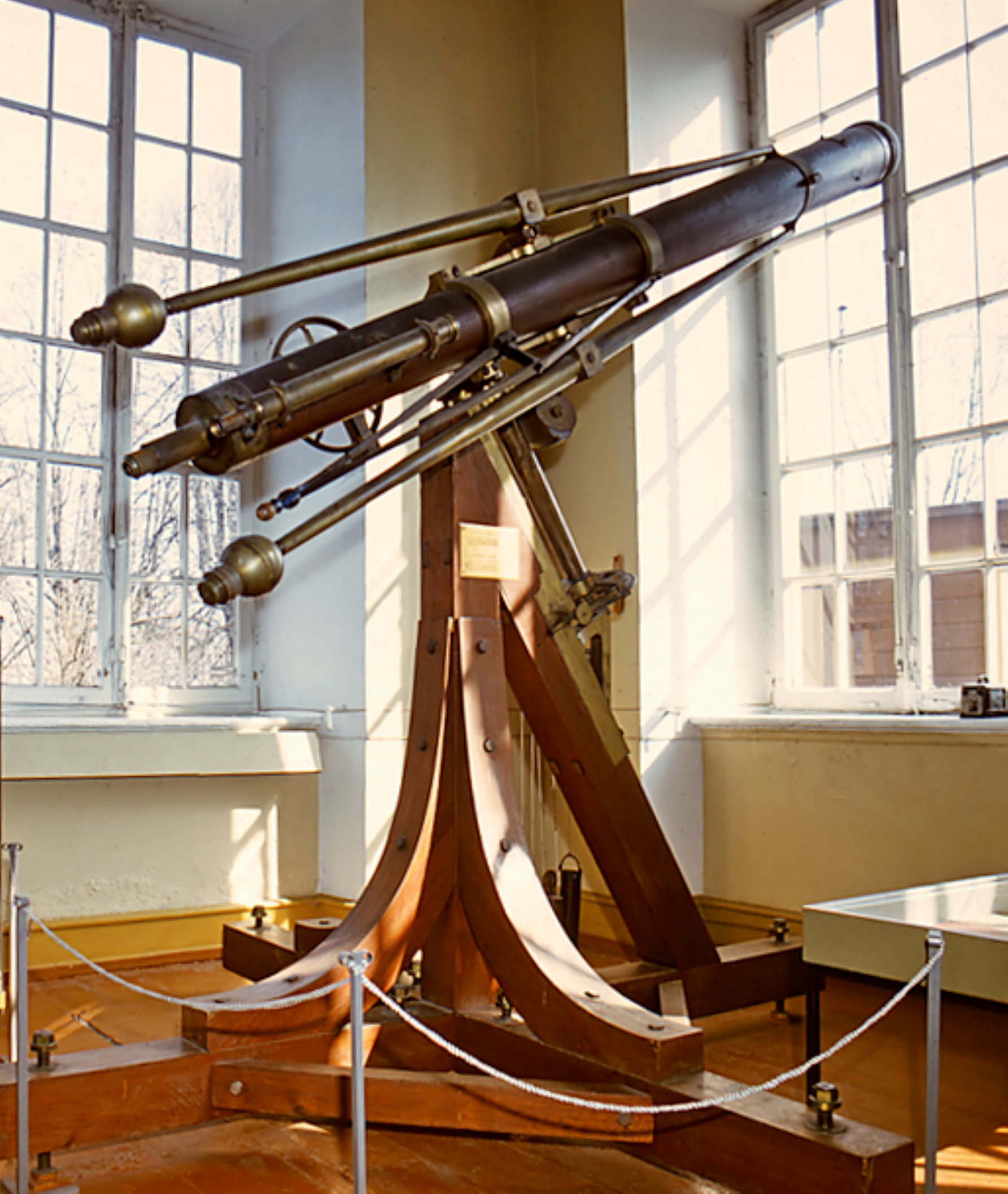}
\caption{\footnotesize Wilhelm Struve's 24-cm refractor, Tartu. This was used in the 1830s to measure one of the first stellar parallaxes, that of the bright star Vega (courtesy Tartu Observatory, Estonia).
\label{fig:struve-refractor}
}
\end{figure}

Wilhelm Struve had selected the bright, high proper motion star Vega for study. At his disposal in Dorpat was a twenty-four centimeter aperture refractor, manufactured by the German physicist and craftsman Joseph Fraunhofer, and the largest instrument of its kind in the world (Figure~\ref{fig:struve-refractor}). Equipped with a `filar micrometer', long used for measuring separations of double stars, two tiny parallel wires or threads, often of fine but immensely strong spider silk, could be moved by the observer using a screw mechanism \citep{1971mckeon}. The slowly changing separations between the target star and a nearby comparison star could be tracked.

Struve's results from seventeen observations starting in 1835 were announced two years later, giving a parallax of one-eighth of a second of arc, close to the present value. But since there had been a long history of fallacious claims to the measurement of parallax, others remained sceptical, and Struve continued his measurements until, in 1840, he gave the results from nearly a hundred observations \citep{1922Obs....45..341J}.

Friedrich Bessel is generally credited as being the first to publish a reliable parallax, spurred on in his measurements by correspondence with Struve and the latter's preliminary result for Vega \citep{1985Ap&SS.110...11F,1988IAUS..133..119D}. From observations made between 1837--38, Bessel carefully tracked the detailed path of the fast-moving binary star known as 61~Cygni, using the heliometer at K\"onigsberg (now Kaliningrad), also manufactured by Fraunhofer. The heliometer had originally been designed to measure the Sun's angular diameter, and hence the name. Its sixteen centimeter diameter refractor lens had been sliced in half, each segment mounted side-by-side, so forming a pair of images which could be adjusted laterally by turning a thumbscrew. Bessel used it to follow the slowly changing angles between his chosen target and a comparison star close by on the sky. Careful monitoring over the course of a year would show a varying separation if the accuracies were sufficient to discern the parallax wobble of the nearby binary. 

In Alan Hirshfeld's readable account of this protracted race \citep{2001hirshfeld}, he describes Bessel's precision instrument as {\it ``almost painfully beautiful: a copper-shaded, mahogany-veneer tube; burnished knobs, gears, and wheels; and a wooden equatorial mount that descended to Earth through a complex of gracefully splayed struts and stout beams.''} To guarantee stability {\it ``the central part of the [telescope] tower's base was filled with five feet of masonry. Atop this were slabs of sandstone and a layer of timbers. Bolted to the timbers were a series of iron-reinforced beams that rose to the upper reaches of the tower and supported the platform on which the heliometer rested.''}

It was an excellent piece of engineering, and with it pointed to the heavens the first star parallax was measured: in 1838, Bessel announced that 61~Cygni had a parallax of 0.314~seconds of arc, placing it at a distance of three parsecs, or ten light-years. What convinced others that a star distance had been measured for the first time was the match between theory and the expected pattern of separations as the Earth moved in its annual orbit around the Sun.

Hot on Bessel's heels was the work of Thomas Henderson, first Astronomer Royal for Scotland, who published a parallax for the nearby star Alpha Centauri in 1839, derived from observations made even earlier in 1832--33 at the Cape of Good Hope \citep{1839henderson}. Although the star is particularly close to the Sun, and its parallax angle therefore amongst the very largest of all stars in the sky, it is only observable from southern latitudes. And with the exception of occasional southern expeditions, such as Halley's and Maskelyne's to Saint Helena, and Abb\'e Nicolas Louis de Lacaille's catalogue of more than ten thousand stars observed from the Cape of Good Hope in the 1750s, the southern skies had received but scant attention. The situation was addressed by England's Board of Longitude which set up a dedicated observatory at the Cape under its first director, the Reverend Fearon Fallows, whom Henderson replaced in 1832 \citep{1913hdro.book.....G,1967VA......9..265E,1988uchs.book.....E}. 

Henderson returned to England barely a year later, dissatisfied with working conditions at the Cape. But included amongst his observations, made with an ordinary mural circle and yet to be analysed, were a series of careful measurements of Alpha Centauri. The star was bright, with a large proper motion, and also one component of a binary with a large separation. It thereby handsomely fulfilled all three of Wilhelm Struve's criteria of likely proximity. 

The announcements of Bessel and Struve, and the star's probable proximity, prompted him to re-examine his own observations from which he duly determined its parallax. Still today, the binary pair of Alpha Centauri, and their fainter companion Proxima Centauri, remain the nearest known stars to our Sun. Pin-pointed from the Hipparcos space measurements, Alpha Centauri has a parallax of 0.742~seconds of arc, which corresponds to a distance of 1.35~parsecs, or 4.396~light-years---just over forty million million kilometers. 

What had at last come together was the understanding that star distances could be measured using the Earth's motion around the Sun, and the pin-pointing of those most promising to measure on account of their likely proximity. Improvements in telescope size, quality, and accuracy, inspired and drove the relentless pursuit.

These first parallax measurements provided the very first rigorous determination of the distances to the stars.  The confirmation that they lay at very great, yet not infinite, distances represented a turning point in the understanding of the Universe. The moment when distances to the stars, and the enormous scale of space, were suddenly and unambiguously revealed must rank as one of the most pivotal in the entire history of science.

John Herschel, President of the Royal Astronomical Society at the time, congratulated members of the society that they had \citep[quoted by][]{1997hoskin} {\it ``lived to see the day when the sounding line in the universe of stars had at last touched bottom.''} In awarding the society's gold medal to Friedrich Bessel in 1841, he described it as {\it ``the greatest and most glorious triumph which practical astronomy has ever witnessed.''}

The refractor used by Struve to measure the parallax of Vega still resides in the museum of the Old Observatory in Tartu, Estonia. Bessel's heliometer, along with the observatory and city of K\"onigsberg, was destroyed in the war-time ravages of 1944--45.

\section{Developments 1850--1980}

Over the period of three hundred years leading up to the detection of the first parallax in 1838, the measurement of star positions had actually followed two somewhat separate branches. The first of these concentrated on the measurement of parallax, exemplified by the pioneering works of Bradley and Bessel.

In parallel were the much larger sky surveys, like those of Flamsteed in the early 1700s at Greenwich, and Lalande in the early 1800s in Paris. For these, the very highest accuracy of individual measurements was sacrificed, and parallaxes were not part of the design. The goal was rather the charting of large numbers of star positions and motions, the motivation being a better understanding of their distribution and their motions through space.

Over the last hundred and fifty years or so, these two branches of star measurements have really split more convincingly into three: small numbers of stars measured with the highest relative accuracy to fix more parallax distances; others spread over the sky and measured with a very good absolute accuracy to give an overall stellar reference frame; and large surveys aimed at elucidating the structure and properties of our Galaxy from the distribution and motion of the stars.

\subsection{Parallax Measurements}

The first of these measurement branches focused on a concerted effort to determine more, and more accurate, parallax distances.  In the years following the first success of Bessel, initial excitement at the prospect of staking out the space distribution of many more stars was overtaken by the bleak realisation that the majority of bright stars lay at colossal distances that still could not be discerned. Observers had to continue to select target stars fastidiously with the best possible prospects of being nearby, while attention still had to be lavished on a relatively small number of candidates. The measurements remained delicate and time consuming. The highest instrument qualities, meticulous checks for any possible errors, and multiple observations throughout the year were all mandatory. 

Visual observations using heliometers continued to dominate until the dawn of the twentieth century. A copy of Joseph Fraunhofer's K\"onigsberg heliometer was installed in Bonn in 1848, and a still larger instrument delivered to Wilhelm Struve's group at the imperial Russian observatory in Pulkovo. Others were procured by observatories at Oxford, Stuttgart, Leipzig, G\"ottingen, and Bamberg in Europe, with the largest such instrument ever made, eight and a half inches in aperture and ten feet long, installed at the Kuffner observatory in Vienna in 1896. David Gill began a heliometer programme in the southern hemisphere at the Cape of Good Hope,  and the first in America was started by W.\,Lewis Elkin at Yale in 1885.

Slowly the number of star distances grew. But progress remained painfully sluggish, and lengthy discussions of the errors reinforced the continuing very great difficulty of the task \citep{1956VA......2.1018J}. Indeed to some it appeared that the era of star parallax measurements was already effectively over; astronomers again, in the words of Hirshfeld, {\it ``defeated by the sheer immensity of the realm they were attempting to chart''.}

What came to the rescue was the new medium of photography. The earliest commercially viable photographic process, daguerrotype, was used by Harvard astronomers J.A.\,Whipple and William Cranch Bond to capture the first photographic image of the bright star Vega in July 1850. More efficient photographic processes appeared, and early celestial astrophotography by amateur Warren De la Rue in England was followed by the first photographic parallaxes by Charles Pritchard at Oxford in 1886. 

Jacobus Kapteyn in Groningen published a list of just 58 parallaxes in 1901. Meridian circles at Leiden and Heidelberg, and photographic plates from Pulkovo and Cambridge, upped the total to 365 by 1910. Yet Kapteyn remained far from satisfied: {\it ``Up to the present and for obvious reasons, parallax observers have devoted their labours exclusively to the bright and swiftly moving stars. In our opinion the time has come for a change of tactics. We need the average parallax of the faint stars and of those with moderate and small proper motion as sorely as the rest.''} His urgent plea was to {\it ``extend the investigations into the arrangement of the stars in space.''}

A new era in photographic parallax determinations was duly opened up by Frank Schlesinger (1871--1943). Astronomy was developing on many fronts, and knowledge of stellar distances became of pressing importance. Schlesinger was born in New York, and his PhD at Columbia University had made use of an unusual benefaction: in 1890, the university had received from the pioneering amateur astrophotographer Lewis Morris Rutherfurd more than a thousand photographic plates of the Sun, Moon, planets and stars taken between 1858 and 1877. Acquired with a thirteen-inch refractor, a particular type of telescope which uses lenses rather than mirrors to focus the starlight, Schlesinger's experience with the plates convinced him that with a high quality telescope of considerable focal length, parallaxes could be determined more economically, more conveniently, and more accurately than by any other method \citep{1927MNRAS..87..506S}.

\begin{figure}[t]
\centering
\includegraphics[width=0.45\linewidth]{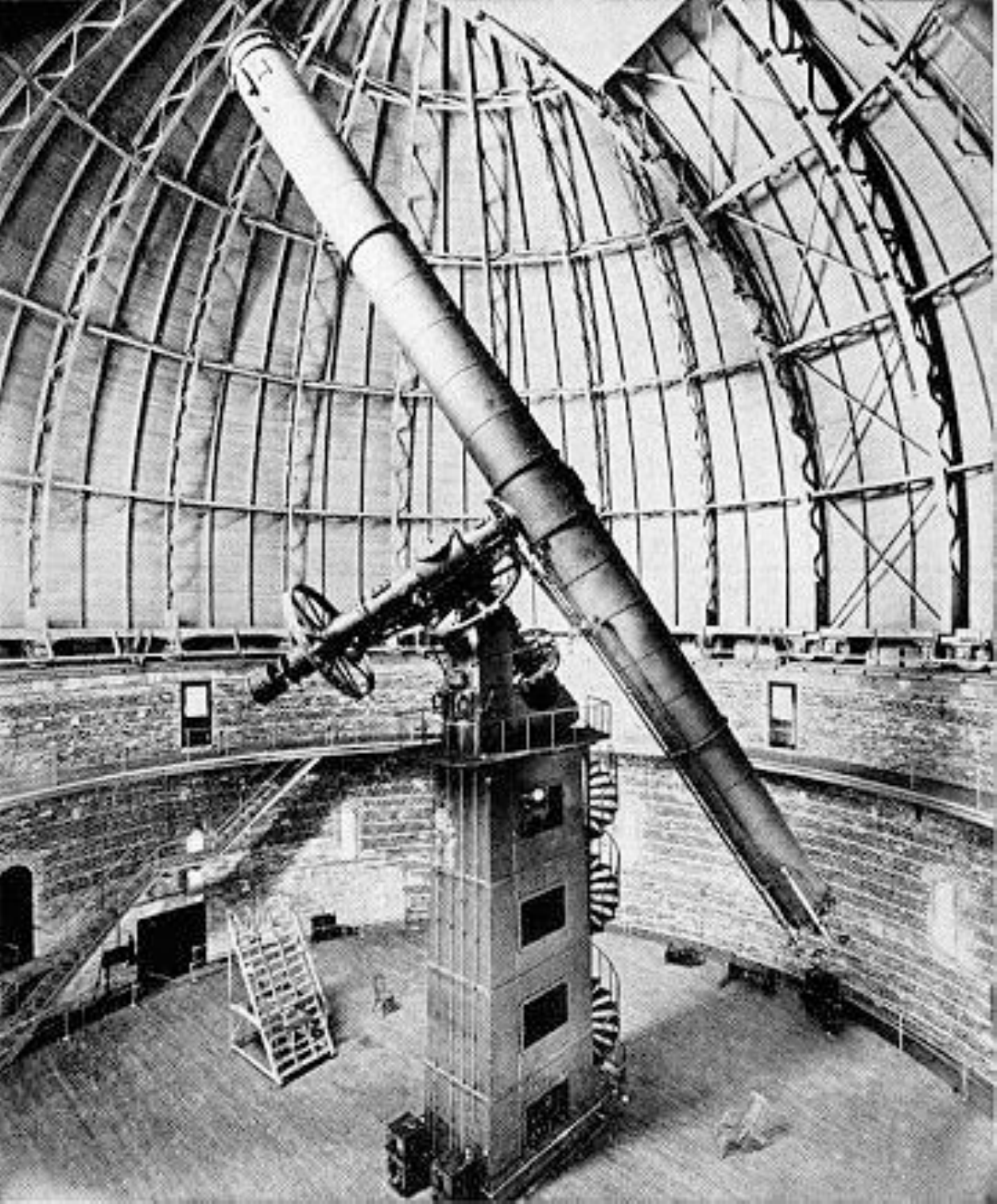}
\caption{\footnotesize The Yerkes 102~cm (40~inch) refractor in 1897. Built by the master optician Alvan Clark, the telescope was exhibited at the 1893 World's Columbian Exposition in Chicago before being installed at the Yerkes Observatory (courtesy Wikimedia Commons).
\label{fig:yerkes-refractor}
}
\end{figure}

So it was that at the Yerkes observatory in Wisconsin in 1903, Schlesinger started a parallax programme using their recently completed forty-inch refractor (Figure~\ref{fig:yerkes-refractor}). This giraffe of a telescope, which remains the largest refractor in the world, was designed around a very long focal length to provide the highest magnification of a small carefully-chosen region of the sky, the easier to discern the tiny parallax wobble.  Measurements under his direction started at the observatories of Allegheny in Pennsylvania, where he served as director from 1905 to 1920, and were continued by his successors at the observatories of Yerkes in Chicago, Van Vleck in Connecticut, and McCormick in Virginia. His classic papers appeared in print in 1910 and 1911, detailing the results for just twenty eight stars. 

Within the next decade, such was his influence, and such was the importance of the task, that eight observatories had made parallax determinations a prominent part of their astronomy programmes \citep{1956VA......2.1023H}. 

His first task as director of Yale university observatory, a position he held between 1920 and 1941, was to plan a new telescope to further the onslaught. A new twenty six-inch photographic refractor of thirty six feet focal length was designed. This time it was destined for the southern hemisphere, to Johannesburg, where it would carry out for the southern skies a programme similar to that at Allegheny for the north. Schlesinger went to Johannesburg in 1924 to supervise the observatory construction, and its subsequent dedication by the Prince of Wales. At the time of his death twenty years later, a remarkable fifty thousand plates had been exposed, and shipped back to Yale university in New Haven for measurement. From this mountain of glass, a further sixteen hundred precious star distances were distilled. Moved to Australia in 1952 due to deteriorating sky conditions, the telescope was destroyed by a fierce forest fire in January 2003---a tragic ending for an instrument which had pinned down the distances of so many of the brightest stars.

In 1924 Schlesinger published his {\it General Catalogue of Stellar Parallaxes}, advancing the total known to just short of two thousand, and extending the frail stellar distance network out to a few tens of light-years. His life's work brought him the gold medal of the Royal Astronomical Society in 1927 and the Bruce medal, another of the highest honours in the field of astronomy, in 1929. 

For almost a century thereafter, parallax determinations were led by American astronomers. It was said of his methods that they were {\it ``basic and complete, and that no major improvements are possible.''} Grand praise, and no great surprise therefore that almost all other parallax programmes of the same era would follow his approach. Outside the United States, Sir Frank Watson Dyson, Astronomer Royal from 1910 to 1933, published twelve years of parallax observations from Greenwich in 1925. His successor as Astronomer Royal, Sir Harold Spencer Jones (1890--1960), published a number of parallaxes of southern hemisphere stars from observations acquired at the Royal Observatory established at the Cape of Good Hope. At a time when many astronomers were moving to newer fields of astrophysics, a few still dedicated their careers to astrometric measurements of the very highest calibre.

In astronomy it often happens that some individual will take the initiative, and rise to the challenge, of making a compilation of all the different work going on around the world in a particular field. With various observatories contributing more distances, often of different quality, and sometimes duplicating attempts at measuring the same star with different instruments, a critical compilation of parallaxes was badly needed. Louise Freeland Jenkins at Yale stepped in to fill a much-needed gap. She brought out a new edition of Schlesinger's {\it General Catalogue of Trigonometric Stellar Parallaxes\/} in 1952, with distances for just under six thousand stars based on photographic determinations of the Schlesinger era \citep{1952QB813.J45......}. A supplement in 1963 raised the total to nearly six and a half thousand \citep{1963gcts.book.....J}. 

A further update appeared in 1995. The {\it Yale Trigonometric Parallax Catalogue\/} of just over eight thousand stars was pieced together by Yale astronomer William van Altena \citep{1995gcts.book.....V}. It was the catalogue that the world's astronomers consulted near the end of the second millennium if they wanted to know the distance to a star. It was also to be the final collection of ground-based parallaxes before those from the European Space Agency's Hipparcos satellite. 

At the time of the push to space, the total number of known star distances was certainly respectable, and had been terribly hard won. But even for the population of stars within our solar neighbourhood it was a paltry sampling, let alone viewed in the context of the hundred billion or more stars in our Galaxy as a whole. Crucial and niggling were the plethora of discrepancies and errors arising from the shimmering atmosphere. Accuracies were supposedly around one hundredth of a second of arc, but in reality were often much poorer. This made it difficult for astronomers to rely on published values, and dangerous to draw wide-reaching scientific conclusions. A new approach to measuring distances was sorely needed.

\subsection{The Stellar Reference Frame}

A second measurement branch was devouring enormous efforts, in parallel with the work on parallax, to set up the best possible stellar reference frame---to measure and list the positions of a number of agreed reference stars ranged across the entire sky. 

To determine a chosen star's distance, repeated positions measured with respect to some other star nearby on the sky would hopefully reveal its parallax motion over the course of a year, but the position of the reference star itself was quite irrelevant. A celestial reference frame demanded, in contrast, a network of precise positions of stars over the entire sky---a set of agreed reference beacons, with positions and motions well nailed down. Hipparchus and Ulugh Beg, Tycho, Flamsteed and Bradley had typified the earliest efforts to establish a stellar reference system across the celestial sphere. 

By the second half of the nineteenth century, a multitude of studies clamoured for a much improved grid of astral trig points. It was needed as a reference frame for the much fainter star surveys starting up to probe the Galaxy's structure, and for studies of the motions of the planets and of the rotation of the Earth. These needs turned to meridian circle instruments to give the best positions for a relatively small numbers of stars. The problem was one that Hipparcos would be well set-up to solve properly later on: that of linking together observations made at different geographic locations and at different times. The reference frame demanded positions of the stars, linked through to the planets, the Moon, and the Sun. A perfidious complication was the fact that the measurement platform, the Earth itself, was slowly `wobbling' due to effects referred to as precession, nutation, and short-term and unpredictable polar motion. 

In Germany, a sequence of whole-sky star catalogues, named the FK~series after the German Fundamental Katalog, began with the work of Arthur von Auwers (1838--1915) in the late 1870s and early 1880s. Their work was to dominate the field for over a century, although a parallel American effort started with Simon Newcomb's accurate charting of just over a thousand stars in 1899, and continued with Benjamin Boss's influential General Catalogue of 1937 \citep{1985CeMec..36..207F}. 

Auwers had started out on his own career at K\"onigsberg, using Bessel's original heliometer. He made his own measurements of a small number of parallaxes, and established the orbital motion of the binary companion of Sirius based on many thousands of meridian circle observations taken over six years. There were, however, no nearby suitable comparison stars for Sirius, and his experiences in constructing a reference system based upon earlier observations led to the catalogue construction work which would dominate the rest of his life. Auwers began by returning to the very accurate observations made by James Bradley over the years 1750--62, comparing them with more modern observations to determine star motions. This piece of work alone would occupy him from 1866 for a further ten years. 

By successive steps, Auwers established a system of just thirty six benchmark stars, with longitudes across the sky fully consistent with each other. Their origin was set by Bradley's observations of the Sun a century before. Into this he folded other observations, of Bradley's own zenith sector measurements acquired from Greenwich, and others by Nevil Maskelyne around the 1760s and by Stephen Groombridge around 1810. The result was a reference system across the sky of just over three thousand stars. All were reobserved from Greenwich around 1865, to give the most accurate motions of stars to date, pinned down from the grand lever arm of a century and a half of meticulous observation. These motions would form the basis of many pioneering researches into star movements carried out over many decades, including Simon Newcomb's revision of the Earth's wobbling motion, and Jacobus Kapteyn's investigations into the rotating Galaxy. The resulting catalogue was published by the Saint Petersburg Academy of Sciences in 1888. In a later collaboration with David Gill in 1889 to refine the distance to the Sun, Auwers provided observational skills much needed by Gill, which the latter acknowledged with the words at the start of this chapter.

These `fundamental' catalogues, it should be stressed, charted only a rather small number of reference stars, dictated by the brightness of stars which could be observed by the meridian circles of the day. They gave only one star every six degrees or so on the sky, or just a handful across the whole of Europe if thought of as a mapping of Earth. Successive catalogues added more observations, and slowly yielded a better grid, although rejecting inferior observations also whittled down the number of quality reference stars. More could be interpolated from meridian circle or photographic observations, but inherent distortions would ultimately rest on the quality of the primary grid.

The final catalogues in the series were prepared at the Astronomisches Rechen-Institut in Heidelberg: the FK4 led by August Kopff and Walter Fricke was published in 1963 \citep{1963VeARI..10....1F}, and the FK5 after a quarter of a century devoted to its upgrade, led by Walter Fricke \citep{1988Obs...108..251M} and published in 1988 \citep{1988VeARI..32....1F}. The work required to create these catalogues extended over many years of careful observation and critical analysis. The FK5 catalogue was the culmination of a compilation of about 260~individual catalogues, observed mostly with meridian circles and some astrolabes. Like the FK4 it contained just 1535 stars. But the scientific importance of these catalogues was nevertheless substantial: they alone provided the reference grid into which the positions of very much fainter star images, captured on photographic plates in their hundreds of thousands during the early 1990s, and in their tens of millions in the later years of the 20th century, could be interpolated.

While individual star positions in these reference catalogues reached accuracies of several hundredths of a second of arc, and notwithstanding the massive effort and observations invested, evidence still suggested that there were significant hidden errors depending on their sky position. Years before, Kapteyn said in 1922 \citep{1922BAN.....1...69K}: {\it ``I know of no more depressing thing in the whole domain of astronomy, than to pass from the consideration of the accidental errors of our star places to that of their systematic errors.  Whereas many of our meridian instruments are so perfect that by a single observation they determine the coordinates of an equatorial star with a probable error not exceeding two or three tenths of a second of arc, the best result to be obtained from a thousand observations at all of our best observatories together may have a real error of half a second of arc and more.''}

\begin{figure}[t]
\centering
\includegraphics[width=0.35\linewidth]{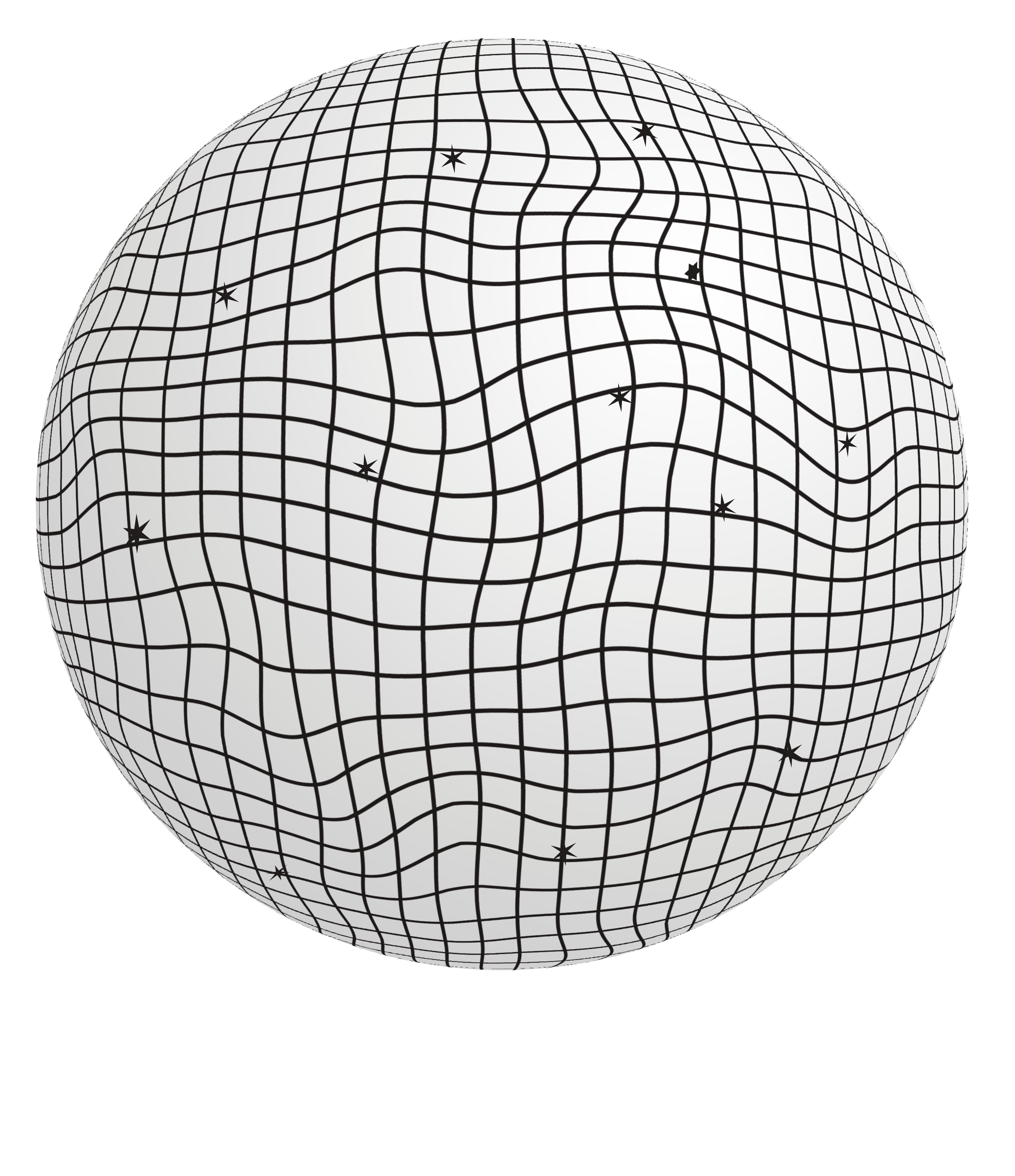}
\hspace{10pt}
\includegraphics[width=0.55\linewidth]{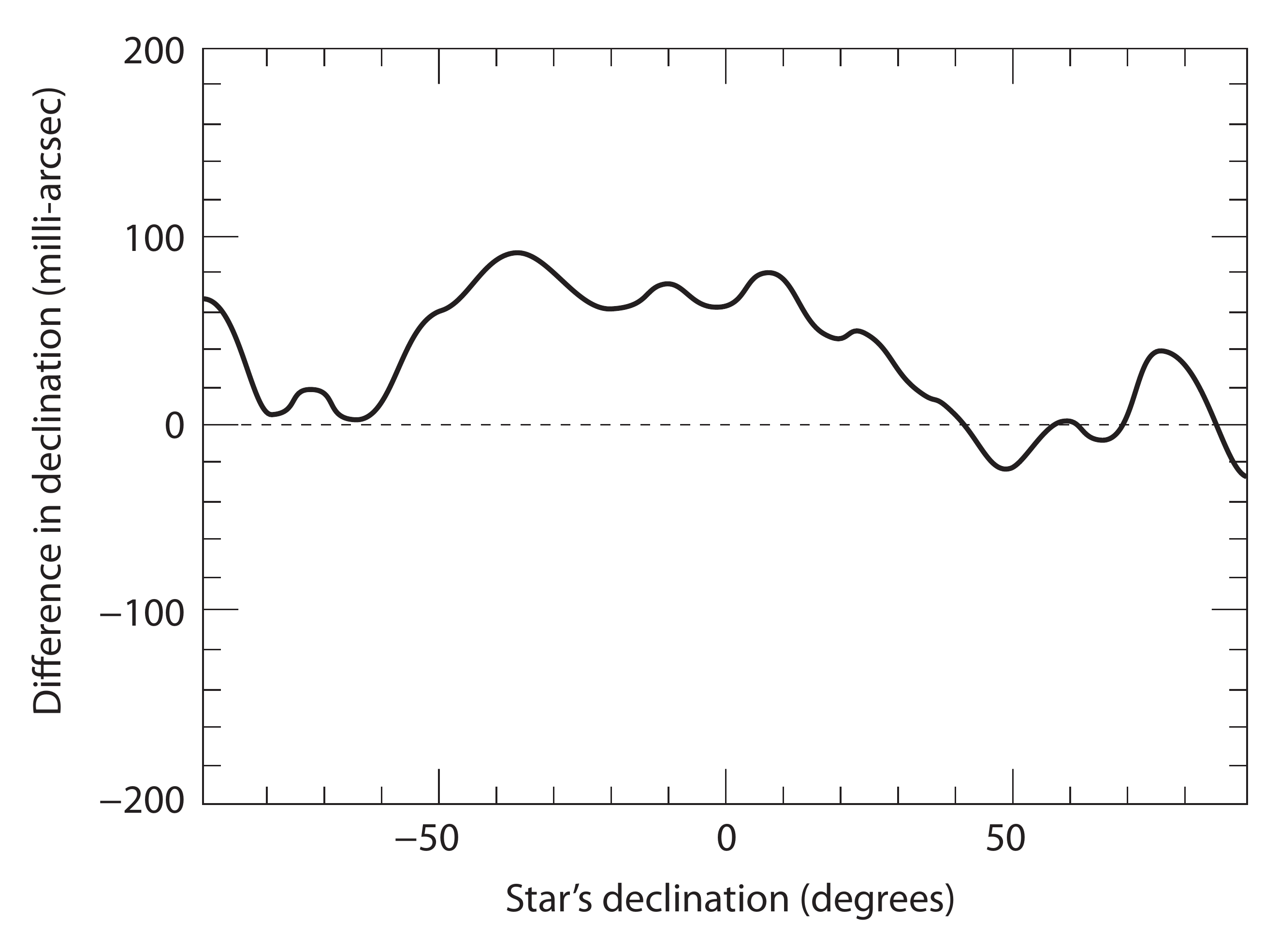}
\caption{\footnotesize (a, left) Schematic of star positions on the sky within a warped reference system. (b, right) example of the systematic differences between star positions measured by the Hipparcos satellite and those from the best ground-based positional catalogue, the FK5, at the comparison epoch 1968.5. This example \citep[adapted from][]{2002A&A...387.1123S} shows the positional difference in declination, as a function of the star's declination on the sky. Compared with the accuracy of the Hipparcos positions, of around 1~milli-arcsec, the warps in the FK5 catalogue reach 100~milli-arcsec or more.
\label{fig:distortions}
}
\end{figure}

Like ancient maps of Earth, the star charts were topologically correct, but stretched and squeezed over the sky in ways that could be guessed but not fully fathomed, hidden errors which proved impossible to track down and remove (Figure~\ref{fig:distortions}a). They were only fully apparent once the Hipparcos space results were published  (Figure~\ref{fig:distortions}b).

Meridian circles remained the instrument of choice for the highest accuracy surveys until the late twentieth century. Most were phased out after the Hipparcos catalogue was published in 1997, but the automatic 18\,cm aperture Carlsberg meridian telescope is one of a few exceptions. It was moved to La~Palma in 1984, refurbished with a CCD detector in 1998, and continues to operate remotely, turning out more than a hundred thousand star observations each night, their positions locked into the Hipparcos grid \citep{2001AN....322..347E}.

The FK5 catalogue of 1535 stars was the final word on the celestial reference frame before the launch of Hipparcos. It was the state-of-the-art in star charting until the space-based positions appeared. But it was far too sparse, and inaccurate, to satisfy modern needs. Like the parallax catalogues, it was impossible to rely on published positions. This was not a good situation for a field so basic. As for the star distances, a new approach to the measurement of a celestial reference frame was required.

\subsection{Large-Scale Surveys}

The third measurement branch of relevance to astrometry over the last century is represented by the large-scale photographic surveys. This branch traces its roots to the early 1600s, when Galileo used the newly-invented telescope to observe the Milky Way, and found that it could be resolved into innumerable faint stars. By the mid-eighteenth century astronomer Thomas Wright (1711--1786) had described the Milky Way as a flattened disk of stars in which the Sun is itself confined, {\it ``an optical effect due to our immersion in what locally approximates to a flat layer of stars.''}

Philosopher Immanuel Kant (1724--1804) developed these ideas, and also postulated the existence of other `island universes' distributed throughout space at enormous distances. William Herschel counted the number of stars in different sky regions to deduce the relative dimensions of our Galaxy. These gave valuable insights, but with conclusions founded on the crucial but incorrect assumption that all stars had the same absolute brightness. It was nevertheless becoming clear that large stellar surveys could have much to say about our Galaxy's basic properties such as its size and its shape.

Many large and enormously influential surveys have been made over the last hundred and fifty years. Important amongst the earliest were the huge three-part `Durchmusterung', named for the German for survey, a word capturing the grandeur of the enterprise. The first two parts were the last of the great star maps to be made visually, pre-dating the use of photography---assistants recorded the positions and magnitudes of stars as the Earth spun and the sky drifted across the fixed telescope field surveying successive latitude zones. The series started with the northern sky surveyed from Bonn by Friedrich Argelander and Eduard Sch\"onfeld. Published between 1852 and 1859, this gave the positions of more than 324\,000 stars of the northern the sky. The extension southwards was surveyed from C\'ordoba in Argentina by John Thome starting in 1892.

The new medium of photography had burst onto the astronomical scene in the late 1800s. Hand-in-hand with the meridian circles giving the highest accuracy reference grid for the brightest stars, photography was to dominate surveys of the skies for the next century. The switch to photography also represented a change in methodology: until then, position measurements had been made by eye, then transcribed to make a star chart. With photography, a chart of the sky was captured directly, and the positions of the stars deduced from them.

Amongst the earliest of these was the southward extension of the Bonn and C\'ordoba Durchmusterung, covering the southernmost skies from the Cape of Good Hope. The results of the work, led by Sir David Gill (1843--1914) and influential Dutch astronomer Jacobus Kapteyn, were published around the turn of the century. Positions were around one second of arc, limited by the twin barriers of atmospheric turbulence and photographic plate quality. The vast Durchmusterungen triptych was only eventually transcribed to computer form in a fifteen-year effort around the 1980s.

These first truly large-scale surveys provided the foundations on which many later investigations would build their own views of the changing positions of the stars. Thereafter new and deeper surveys from many different observatories around the world contributed to the growing edifice.  

Photography allowed the positions of stars to be measured in terrific numbers. With the large telescopes and long exposures of the later 1900s, deep sky images several degrees in extent could yield thousands or millions of star images per plate. The technique was straightforward in principle: exposed at the focus of a telescope tracking the apparent motion of the celestial sky, the plates provided images of stars in large numbers. Their positions on the plates could then be measured and recorded, duly transformed to provide immense catalogues of star positions. In practice, good images require excellent high-altitude observing sites, excellent telescope optics, and accurate and smooth drive mechanisms to track the rotating sky. But they could never eliminate the straightjacket imposed by the atmosphere.

Photographic plates store well for decades, and astronomical libraries and archives across the world preserve a record of how the skies appeared over the past century. As the technology reached its peak in the 1970s and 1980s elaborate and fast automatic measuring machines scanned new and ancient archive plates wholesale. Together they have captured and stored the results in the form of huge digital catalogues of the night sky which will be preserved indefinitely. Yet fundamental distortions due to the telescope optics have always confounded the ultimate accuracies, while the Earth's atmosphere, and the ever-so-slightly dancing images seen through it, still imposes its ever impenetrable barrier. 

Although the highest positional accuracies were therefore sacrificed in favour of quantity of stars, massive sky surveys using photographic plates nevertheless changed the course of astronomy. There were various reasons for this impact. First off, simply counting stars to different brightness limits in different directions of the Galaxy has provided many clues as to its structure and dimensions. The technique is especially powerful when interpreted alongside other knowledge, such as the type or temperature of the stars from spectroscopy.

Measuring the same region of sky over many years or decades is particularly effective at revealing the motions of many stars. Photographic surveys, carefully calibrated and repeated decades later, have turned the early detections of star motions by Halley and others into a large-scale discovery factory on an industrial scale. Repeating exposures of the stars over intervals of months or years has another important spin-off: it has led to the discovery of huge numbers of variable stars, their variability over time encoding clues as to their masses, luminosities, and evolutionary states.  Star colours measured from different filters and photographic emulsions also provide a wealth of indicators such as their temperature and gravity.

Star positions in large numbers allowed astronomers to embark on a new, more quantitative discussion of our Galaxy's structure. In 1904, studying the Cape Photographic Durchmusterung, which he had worked on in collaboration with David Gill, Kapteyn found that the motions of stars were not random, but could be divided into two streams, moving in nearly opposite directions in different parts of the sky---the first hint of the rotation of our Galaxy. In a summary of his life's work published in 1922, Kapteyn described the Galaxy as a lens-shaped island universe in which the density of stars decreased away from its centre. In his model, the Galaxy was held to be some 40\,000 light-years in size, not so far from our present ideas. But, as if clutching at long-held belief that the Earth must occupy some privileged place in the Universe, Kaptyen held that the Sun was relatively close to its centre, at around two thousand light-years.

The size of the Galaxy, and the distance scale within it, became issues of great debate.  It was not easy to infer the structure of the Galaxy from star counts alone, and there were many complications. Great clouds of dense interstellar gas occupy various pockets within our Galaxy's disk, and these block out the more distant light from stars beyond. It's not so different to looking at the night sky covered by thin cloud. With no simple means to identify the gas, seeing only a few stars along a particular sight line might suggest that the Galaxy was only thinly populated by stars in that direction, while the very opposite might be true. Another tricky problem was caused by the growing realisation that stars were of many different types, with hugely varying luminosities and very different types of motion through space. So evident in retrospect, trying to figure out the properties of our Galaxy from an erroneous census was doomed to fail.

So it was that even into the 1920s, the detailed structure of our own Galaxy, and the relationship between it and those that we now know lie far beyond, still remained a puzzle. The uncertainties precipitated an exchange which has gone down as astronomy's Great Debate, which took place on 26~April 1920 in the Smithsonian Museum of Natural History in Washington~DC. Harlow Shapley, from the Mount Wilson observatory, argued that our Sun lay far from the centre of a single Great Galaxy, in which spiral nebulae such as Andromeda were simply part of our own. Heber Curtis, from the Allegheny observatory, disagreed. He held that the Sun was near the centre of a relatively small Galaxy, with the entire Universe composed of many other galaxies somewhat like our own. It was a debate deeply rooted in the uncertainty of the scale of the Universe which had still not been resolved. 

Edwin Hubble's identification of pulsating Cepheid variables in the Andromeda nebula in the mid-1920s confirmed that it was a distant galaxy much like our own, but far beyond. Like brilliant lighthouses pulsing across the depths of space, these standard candles illuminated our understanding of the scale on which the Universe is constructed. Shapley was duly proven more correct about the size of our Galaxy and the Sun's location within it. But Curtis's view that the Universe was composed of many more galaxies, and that `spiral nebulae' were indeed galaxies just like our own, was corroborated. With almost a century's hindsight, the debate is important, in the words of \citet{1982shu}:  {\it ``not only as a historical document, but also as a glimpse into the reasoning processes of eminent scientists engaged in a great controversy for which the evidence on both sides is fragmentary and partly faulty.''}

Schmidt telescopes appeared on the scene in the second half of the twentieth century, and brought their own revolution. Named after their optical designer Bernhard Schmidt, a cleverly-designed `corrector' lens positioned in front of the primary reflecting mirror resulted in strongly reduced image aberrations over unprecedentedly large fields of view of several degrees on a side (specifically, the design allows very fast focal ratios, while controlling coma and astigmatism). This made it possible to observe a substantially larger region of the sky, several times the diameter of the full Moon, in a single exposure. As  a result, Schmidt telescopes contributed a flood of high-quality observations that brought positional astronomy back to the fore. 

\begin{figure}[t]
\centering
\includegraphics[width=0.55\linewidth]{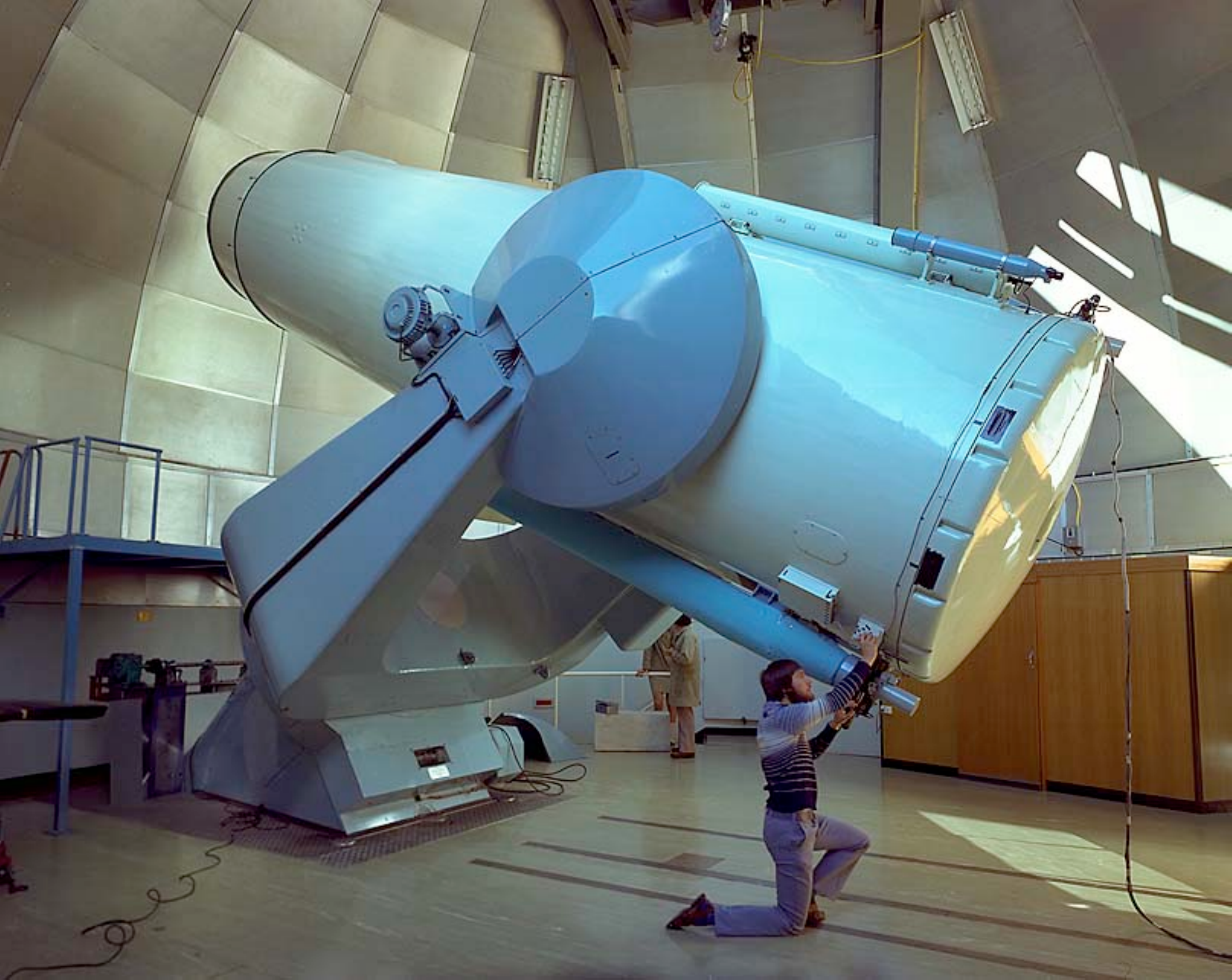}
\caption{\footnotesize The UK 1.2-meter Schmidt telescope at the Anglo--Australian Observatory, Australia (1977), used for large-scale photographic surveys of the southern skies from the 1970s (courtesy Australian Astronomical Observatory and David Malin Images).
\label{fig:schmidt-aao}
}
\end{figure}

Monumental surveys were carried out from Palomar Mountain in California from 1949, in a grand programme funded by a grant from the National Geographic Society to the California Institute of Technology. The southern skies were surveyed by the European Southern Observatory from the mountain top observatory on La~Silla in Chile from 1973, and from the UK's observatory in Australia from about the same time (Figure~\ref{fig:schmidt-aao}). 

The surveys produced thousands of meticulously exposed plates which were themselves reproduced photographically, and circulated in limited editions to the world's astronomical institutes for detailed scrutiny.  Collectively, they comprise hundreds of billions of star images, an archival view of the celestial sky as it will never be seen again. The resulting vast catalogues, of more than a billion stars across the sky, are used for countless astronomical projects, including pointing their way around the sky by the great space observatories, the Hubble Space Telescope amongst them. 

Photographic plate surveys made far in the past---a century or more ago---remain of value to present day astronomy, for a repeat survey today will easily identify the most rapid movers with the largest motions. Catalogues of stellar motions continue to be constructed from various combinations of these photographic plates, using the same technique which allowed Edmond Halley to identify the first stellar motions three hundred years ago.

\subsection{The Carte du Ciel}

In this context, one remarkable project deserves specific mention: the imposingly named {\it Carte du Ciel}, the Map of the Heavens.  It is noteworthy not so much for its profound scientific achievements, but rather for its hugely ambitious scale. This vast and unprecedented international star-mapping project was initiated by ex-naval officer and Paris Observatory director Rear Admiral Am\'ed\'ee Mouchez, in collaboration with Sir David Gill, Her Majesty's Astronomer at the Cape of Good Hope at the time. 

Mouchez had started his career with hydrographic studies of the ocean depths, tides and currents along the coasts of Korea, China and South America and later, during the Franco--Prussian War, led a heroic defence of Le~Havre. Taking the helm at the Paris Observatory, correspondence between Mouchez and Gill led to the {\it ``assembling of a great international conference''}, the Astrographic Congress of more than fifty astronomers held in Paris, on 16~April 1887. Participants included Auwers from Germany, Kapteyn from The Netherlands, Struve from Russia, and William Christie, the Astronomer Royal from England \citep{1912turner}.

The new medium of astronomical photography offered a remarkable possibility to carry out a celestial survey totally unprecedented in the history of astronomy, and astronomers seized the opportunity. The objectives of this first ever international astronomical collaboration on a massive scale were hugely ambitious but would prove to be overwhelming. The idea was to build up and deploy a system of identical telescopes straddling the full range of latitudes on Earth, survey the sky, and build up a monumental star catalogue as a result. According to H.\,H.~Turner's highly-readable description of the project \citep{1912turner}: 
{\it ``The discussions were, to say the least of it, animated. There are no universal rules for conducting public business, and astronomers from one country were not familiar with rules in use elsewhere. It interested Englishmen, for instance, who are accustomed to have resolutions moved by anyone rather than the chairman, to learn that this was by no means a universal rule. On the contrary the chairman of the first conference considered it part of his duties to move all the resolutions. After listening to a discussion, he took it to be his function to summarise the sense of the meeting in a resolution which he put from the chair and in favour of which he held up his own hand. Unfortunately for his success his was sometimes the only hand held up, and the discussion was necessarily resumed.''} Turner considered that the conference was:
{\it ``\ldots a remarkable meeting, the first of its kind in the history of astronomy; and it has shown the way for subsequent gatherings\ldots\ On all of these occasions the French have acted as hosts and have discharged these duties with a cordiality and hospitality that has never failed to impress their colleagues from the most distant parts of the world.''}

The ambitious enterprise had two separate yet connected parts. The first, the Astrographic Catalogue, would photograph the entire sky to 11~magnitude, thereby picking out stars a hundred times fainter than the feeblest seen by the unaided eye. It would provide a plentiful reference catalogue much denser than anything observed by transit instruments.  

Twenty observatories around the world participated, each choosing a strip of sky convenient in latitude. Each would procure the necessary astrograph (a telescope designed specifically for the purpose of astrophotography), suitably equipped and staffed. Then collectively they would expose, for six minutes each, more than twenty thousand glass plates of the night sky. Turner estimated the total weight of these plates at three tons.  

\begin{figure}[t]
\centering
\includegraphics[width=0.45\linewidth]{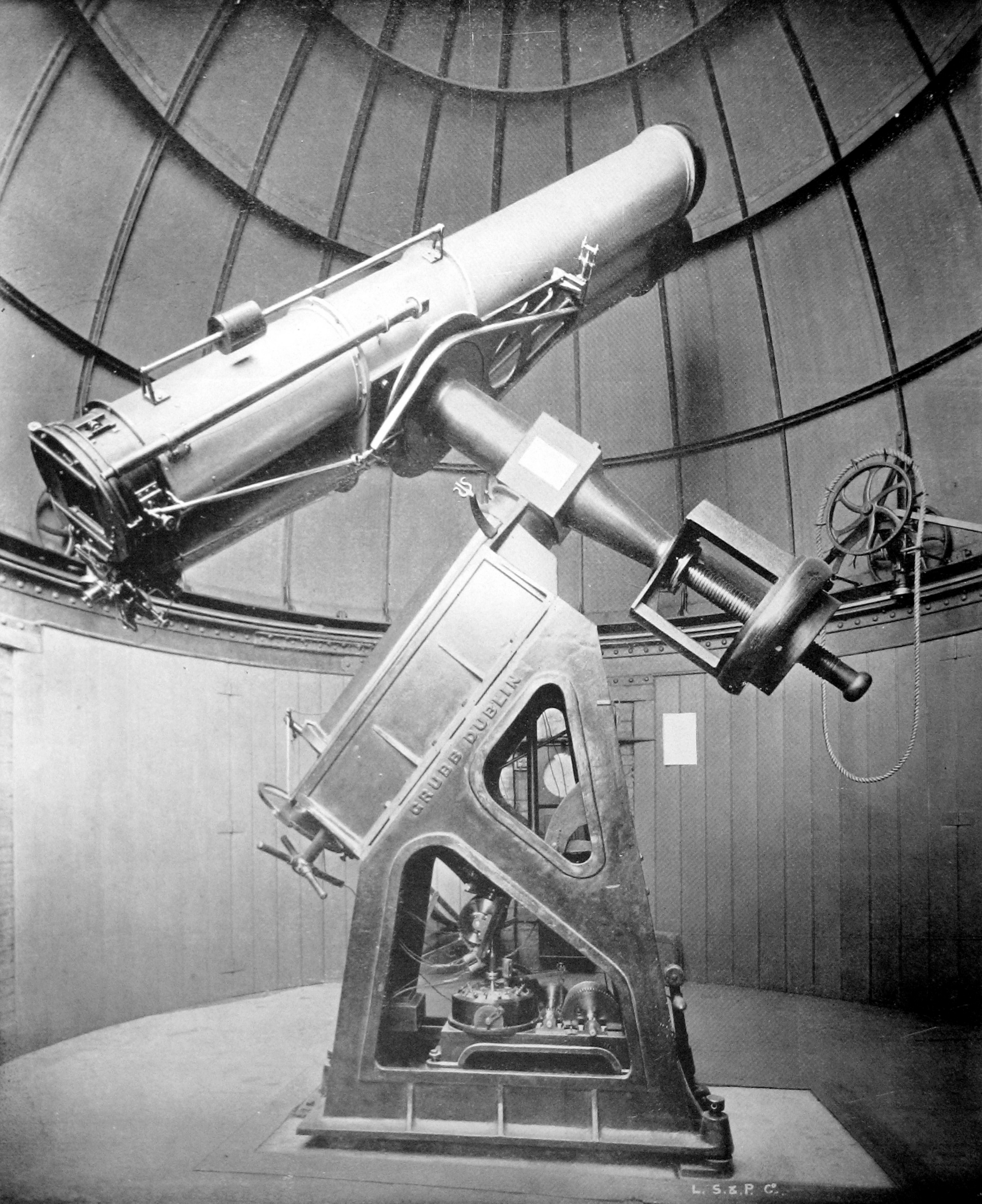}
\caption{\footnotesize The Greenwich astrograph, procured for the international Astrographic Catalogue/Carte du Ciel programme, c1900 (from the Greenwich Zone Astrographic Catalogue, courtesy Graham Dolan).
\label{fig:astrograph-greenwich}
}
\end{figure}

A key agreement, and one essential to the survey uniformity, was to use similar telescopes. Around half of the observatories eventually procured astrographs from the Henry brothers in France, with the others coming from the firm of Howard Grubb in Dublin.  The different observatories were assigned different latitude strips to photograph: Greenwich, the Vatican, Catania, Helsing, Potsdam and Hyderabad would cover the northern sky (Figure~\ref{fig:astrograph-greenwich}). Uccle, Oxford, Paris, Bordeaux, Toulouse, Algiers, San Fernando and Tacuba would span the equatorial regions. C\'ordoba, Perth, Cape of Good Hope, Sydney, and Melbourne would survey the southern skies.

The first plate was taken in August 1891 at the Vatican Observatory. The exposures there, taken by the hands of a single observer, took more than twenty seven years to complete. The very last plate was finally exposed in December 1950 at the Uccle Observatory in Bruxelles.  

The plates were in due course photographed, measured, and the results published in their entirety, providing star positions with an accuracy of about half a second of arc.  In practice, the measurements were a highly protracted affair, with the tasks around the world assigned to willing---and in some cases unwilling---assistants.

Adriaan Blaauw recalls that Pieter van Rhijn (1886--1960), Kapteyn's successor as director of the Astronomical Institute in Groningen and who Blaauw himself knew well, had told him that Kapteyn had numerical computations of star coordinates carried out by prisoners in Groningen. According to Blaauw: {\it ``A number of these tables still exist and are now part of the Kapteyn legacy collection kept in the Groningen University Library where they can be consulted. They are a marvel of neatness and accuracy. The people who made them must have taken great pride in delivering them and one can imagine that it must have given them great satisfaction to contribute in this way to Kapteyn's scientific work.''} Doubts were raised about the role of prisoners at the Kapteyn Legacy Symposium in 2000, there being no written documentation, but Blaauw vouched for the story's pedigree. 

\begin{figure}[t]
\centering
\includegraphics[width=0.5\linewidth]{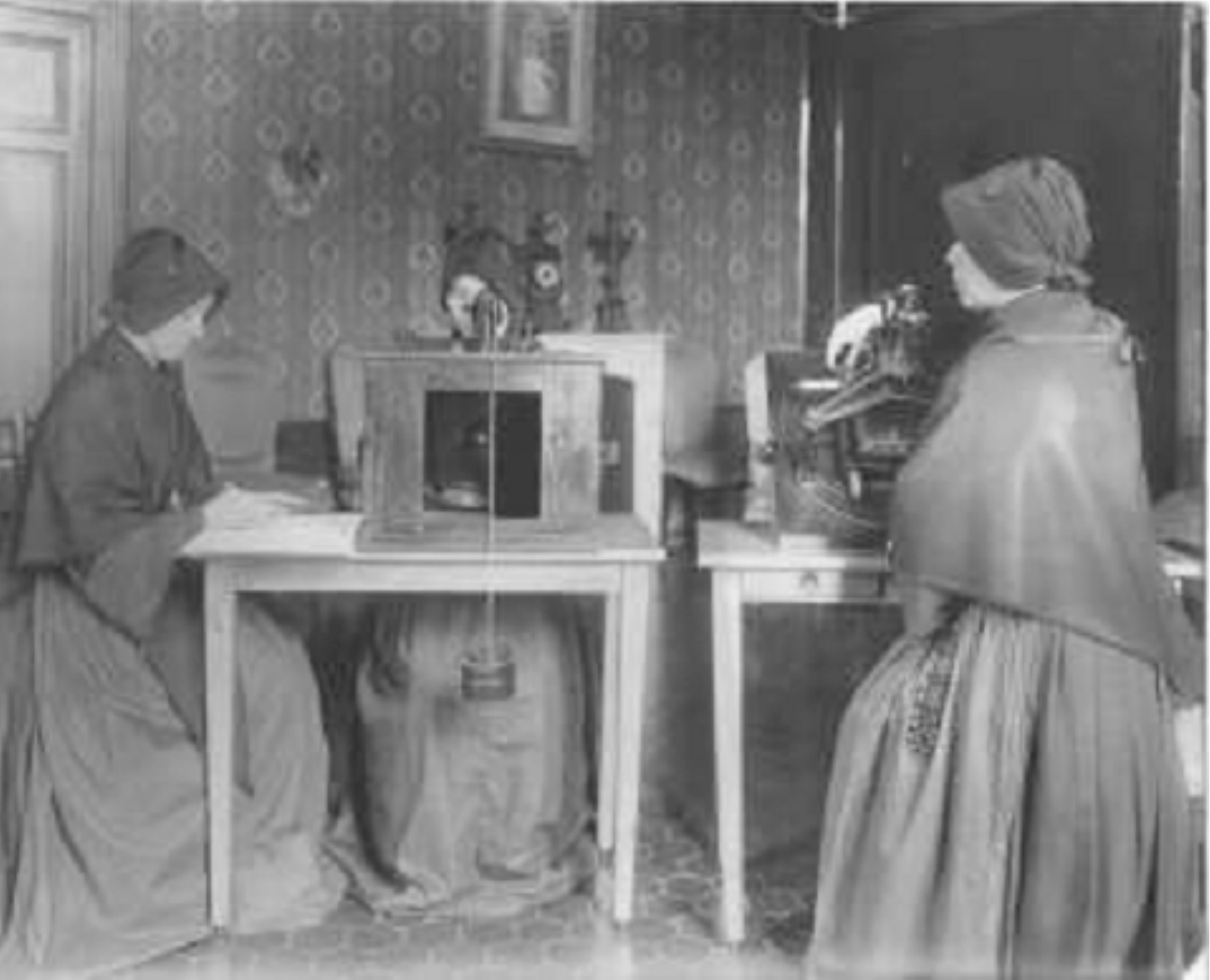}
\caption{\footnotesize Nuns measuring the Vatican Zone Carte du Ciel plates, c1900 (courtesy Torino Observatory).
\label{fig:cdc-nuns}
}
\end{figure}

All measurements of the star images were made by eye, and recorded by hand. In many observatories---Paris, Melbourne,  Perth, Cape, Toulouse and others---twenty or thirty women (the original `computers') were taken on to help with the herculean task. For the Vatican plate collection, archival photographs show nuns from the Congregation of the Child Mary at work measuring the plates (Figure~\ref{fig:cdc-nuns}). Turner commented that {\it ``each observatory has thus to measure about half a million star images\ldots\ These measures took a staff of four or five people at Oxford some ten years or so to complete: and the printing of them another four years.''} In total, nearly five million stars were recorded. Publication of the various parts proceeded from 1902 to 1964, and resulted in a massive two hundred and fifty four printed volumes.

For the second part of the conference goals of 1887, a further set of plates, with longer exposures but minimal overlap, would photograph all stars to 14~magnitude, corresponding to stars a thousand times fainter than those that can be seen with the naked eye. Most of these plates used three exposures of twenty minutes each, displaced to form a small triangle with sides of ten seconds of arc, making it easier to distinguish stars from plate flaws, and to differentiate stars from the more rapidly-moving asteroids. The grand idea was that exposed plates would be reproduced and distributed as a set of charts, the {\it Carte du Ciel}. However, reproduction of the charts, originally to be undertaken using engraved copper plates, proved to be prohibitively expensive, and many zones were either not completed or not properly published. 
  
Despite, or perhaps because of, its vast scale, the project was only ever partially successful, even though many committed individuals had devoted decades of their careers to its success. The {\it Carte du Ciel\/} component was never completed, and the Astrographic Catalogue lay largely ignored for nearly a century. Its star positions were difficult to work with because they were not available in computerised form, and neither were they listed in convenient coordinates.  Some historians of science have classified this vast project as the story of how the best European observatories of the nineteenth century lost their leadership in astronomy by committing vast resources to a somewhat misguided undertaking.  

Long portrayed as an object lesson in over-ambition, languishing lost and forgotten for a century, the Astrographic Catalogue made a remarkable reappearance on the world's astronomical stage around a decade ago. The Hipparcos catalogue positions could be used, in combination with each star's proper motion, to provide a reference frame back at the time when the Astrographic Catalogue plates were taken. So calibrated, they gave the places of all catalogue stars which they occupied in the sky some one hundred years before. Combining those with the satellite positions nearly a century later gave extremely accurate motions for two and a half million stars: the Hipparcos satellite-based Tycho~2 Catalogue, led by the influential Danish astrometrist Erik H{\o}g \citep{2000A&A...355L..27H}. 

Like the ancient catalogue of Hipparchus dusted off and used to reveal star motions by Halley, the Astrographic Catalogue is a remarkable example of an all-but-abandoned project, for whom so many had toiled for so long, waiting patiently to prove its inestimable value generations afterwards.

\subsection{Other Photographic Sky Surveys} 

Numerous other large-scale photographic astrometric sky surveys were carried out in the twentieth century. The following chronology of some of the major developments in twentieth century astrometric surveys is intended only to set the context. A more complete summary, with detailed references, is given in Table~2.3 of \citet{2009aaat.book.....P}.

AGK2: between 1928 and 1931, the sky north of declination $-5^\circ$ was photographed on 1940 glass plates each covering over $5^\circ\times5^\circ$ with two dedicated astrographs located in Bonn and Hamburg, Germany. Two exposures, one of 3~minutes and one of 10~minutes, were made on each plate, and reached about 12\,mag. During the 1930s--1950s the measuring and reduction of the brighter stars were carried out, by hand, resulting in the AGK2 Catalogue.
    
AGK3R and AGK3: after a proposal that the AGK2 Catalogue should be observed again at Hamburg to provide proper motions, an extensive international programme of meridian observations at ten observatories was organised, under IAU Commission~8, to provide a reference star catalogue, AGK3R, which was then used for the reduction of the photographic work carried out at Hamburg between 1956--63. This resulted in the AGK3 Catalogue, containing proper motions for all stars, which was subsequently used as the stellar reference frame in the northern hemisphere.
  
SAO: by the mid-1960s a high density catalogue of star positions was needed for satellite tracking. This was compiled by the Smithsonian Astrophysical Observatory for more than 250\,000 stars. In each declination zone, preference was given to source catalogues with proper motions, namely the Yale Photographic Catalogues in the north, and the Cape Catalogues in the south. The resulting SAO Catalogue was limited by the generally poor quality of the first epoch material in both hemispheres (the AGK3 not yet being available in the north). Not surprisingly, in view of the inhomogeneous source material used in the construction of the SAO, the differences with the later Hipparcos results show various large distortion patterns.

SRS: the success of the AGK3R programme led to plans for a similar campaign in the southern hemisphere, formulated by the International Astronomical Union in 1961. The resulting Southern Reference Star (SRS) Catalogue was constructed from observations made with 13~transit circles, with observations extending from 1961 for about two decades. The International Reference Stars (IRS Catalogue) comprises the combination of the resulting reference stars from both hemispheres, i.e.\ the AGK3R in the north and the SRS in the south.
  
CPC2: to complement the AGK3 in the northern hemisphere, the Second Cape Photographic Catalogue, CPC2, was constructed from 5820~southern hemisphere plates taken with a new astrograph at the Cape Observatory during 1962--1972 (mean epoch 1968), and scanned with the GALAXY plate measuring machine at the Royal Greenwich Observatory, Herstmonceaux. This resulted in a catalogue of 276\,131 stars in the range $6.5-10.5$\,mag. 

Many of these (and other) grand twentieth century photographic surveys have been revitalised by the results of the Hipparcos satellite mission. The new reference system from space can be propagated backwards in time using the measured proper motions, to give an improved reference system for the years that the plates were taken. The improved reference system then gave much better positions for the large numbers of other stars on the plates. This, in turn, has led to vastly improved star motions tracked between the times of the earliest photographic plates a century ago, and the measurements from space made in the last decade of the second millennium.

\subsection{Accurate Solar System Measurements} 

A final mix of curious phenomena showed up in the measurement of the accurate positions of the stars and the planets over the last couple of centuries, bringing us back, in full circle, to the earliest of the Greek studies of the fixed stars and the wandering planets.

Objects in our daily lives are generally not massive enough, or the effects not measurable accurately enough, for Newton's Law to be examined for real flaws or imperfections. But the motions of the planets provide a miraculous laboratory for observing the most delicate touches of gravity. Alongside innumerable other successes of Newtonian gravity was its part in the discovery of the planet Neptune. 

In the middle of the nineteenth century French mathematician Urbain Le~Verrier (1811--1877), working under Fran\c cois Arago at the Paris Observatory, had been making a careful study of the orbit of Uranus \citep{2009lequeux}. There were small but systematic discrepancies between its observed orbit, and that predicted by Newtonian theory---its measured position was consistently off from where theory forecast it should be. Something was wrong. 

Newtonian gravity had proven itself repeatedly and was not the suspect. Le~Verrier was forced to conclude that an undiscovered planet existed out in the far reaches of the solar system, giving erratic tugs at Uranus during its journey around the Sun. He could predict a position for an unknown object which, he believed, must be responsible for disturbing its orbit. Neptune, as it would be called, was duly discovered by Johann Galle and Heinrich d'Arrest, within one degree of his predicted location, on 23~September 1846 \citep{1846MNRAS...7..153G}. It was a triumph for Newtonian gravity, and a sensational result for Le~Verrier, who became director of the Paris Observatory in 1854, following in the footsteps of Cassini and Lalande. A source of debate ever since has been the extent to which John Couch Adams, who had made similar calculations even earlier, should also be credited with Neptune's discovery.

The earliest and most worrying sign that all was not completely well with Newtonian theory was the detailed motion of our innermost planet. Mercury circles the Sun in a tight, bakingly-hot elliptical orbit of just ninety days. Its point of closest approach advances around the Sun by a small amount each year, about one minute of arc, due to various effects, including the gravitational pull of the other planets. 

Le~Verrier noticed that the slowly changing shift could not be explained completely by Newton's laws \citep{1982mpfl.book.....R}. There was a tiny mismatch of a little less than half a second of arc per year, an almost undetectable amount, except for the fact that it rolls up and accumulates with time, to nearly forty three seconds of arc each century. In 1843, inspired by his success with Neptune, Le~Verrier published his interpretation of the mismatch as being due to a hypothetical inner planet, which he named Vulcan. This precipitated a search for the new planet, and a wave of false detections that would follow unabated over the next sixty years. One Edmond Lescarbault was even awarded France's prestigious {\it L\'egion d'honneur\/} for his claimed sighting of the non-existent body.

In 1915, while the searches were in full swing, Albert Einstein published his general theory of relativity. This describes gravity as a basic property of the geometry of space and time, a distortion in their very fabric due to the presence of mass. It superseded Newton's law of universal gravitation as `the' theory of gravity. Mathematicians admire its elegance, and physicists like it because it gives hints as to why this force exists. Mostly the predictions of Newton and Einstein agree. But in certain situations they differ, slightly but significantly, and tests to confirm or repudiate it were eagerly sought. 

The orbit of Mercury was an obvious target. It was Einstein himself who showed that his theory explained exactly the discrepancy, important evidence that he had identified the correct form of the equations describing gravity \citep{1915SPAW...47..831E}. The effect, referred to as perihelion precession, has also been seen for Venus and Earth. In a very close binary pulsar system, discovered in 1974, the effect is a hundred thousand times larger. In all cases, theory and observation are in precise accord. 

Le~Verrier died in 1877 still convinced that he had detected a second planet. Yet while most of the interest in Vulcan evaporated, claims and counter-claims of asteroid transits, and searches for Vulcanoid asteroids orbiting close to the Sun, continue to the present. 

Another test proved to be still more compelling. According to the prescriptions of general relativity, starlight should be deflected by a very tiny but entirely predictable amount as it passes from a distant star close to the limb of the massive Sun on its way to an observer on Earth. The size of the deflection was predicted to be very small, just over one second of arc at the limb of the Sun where the effect would be largest. Barely at the limit of the dancing motion of the atmospheric ripples, it would demand careful measure, and an excellent knowledge of the undeflected star image positions to compare with. 

It would be impossible to measure position shifts of faint stars close to the limb of the brightest object in the entire sky except, perhaps, if they could exploit the exceptional conditions of a total solar eclipse. This was American solar astronomer George Ellery Hale's proposal to Einstein when asked to suggest an appropriate test. A German--USA expedition planned for an eclipse passing over Crimea in 1914 was foiled by the outbreak of war. The first observations of this light bending were eventually made during the total eclipse of 29~May 1919. Astronomer Royal Sir Frank Watson Dyson had identified this as an auspicious celestial alignment because the Sun and Moon would pass in front of the bright Hyades cluster, more bright stars making it easier to detect changes in their position. The undeflected star positions that would later be observable close to the Sun's limb during the eclipse had been observed six months previously by night. 

Arthur Eddington and Edwin Cottingham from Cambridge journeyed to the West African island of Pr\'incipe in the Gulf of Guinea, while Andrew Crommelin and Charles Davidson from the Royal Greenwich Observatory set up their base near the Brazilian town of Sobral---the two observing stations chosen to improve prospects of observing the eclipse in case of poor weather. During the eclipse, as the sky was plunged into darkness, a few bright stars popped into view and remained visible for two or three minutes. This time, their positions would be minutely deflected by the presence of the Sun's huge gravitating mass along the light path from the distant stars behind the Sun to observers on Earth. 

The agreement between the small extra shifts observed on the one hand, and Einstein's theory on the other, was very much at the limit of star measurement accuracies of the time. Confirmation of the predicted bending was duly claimed \citep{1920RSPTA.220..291D}, and widely greeted as spectacular news. It made the front page of major newspapers, making the theory of general relativity world famous, and Einstein himself even more so.  When asked what he would have said had his theory not been proven by the observation, Einstein notoriously replied {\it ``I would have had to pity our dear Lord. The theory is correct all the same.''}

Debate about the quality of these early observations has continued, but the theory itself is now unquestioned. Better measurements for other solar eclipses, including one in June 1973 by Texan astronomers from a desert site near Chinguetti in Mauritania \citep{1976AJ.....81..455J}, sightings of quasars at radio frequencies, gravitational lenses observed in astronomy in the 1980s, gravitational redshift as perfectly accounted for by GPS navigation satellites, and many other subtle manifestations, have confirmed general relativity as our best description of gravity to date.

\section{Developments Since 1980}

\subsection{The Advent of Solid-State Detectors}

In the last 20--30~years, photographic plates have all but disappeared from astronomy, going the way of sextants and quadrants and most meridian circles before them. In their place the CCD, the ultra-sensitive solid-state silicon detectors, of the type used in digital cameras and video recorders (and comparable infrared-sensitive detectors), has taken over the challenge, and has brought with it another revolution in surveying the skies. 

The full-sky surveys of the US~Naval Observatory, notably USNO\,B \citep{2003AJ....125..984M} and UCAC2 \citep{2004AJ....127.3043Z}, and the Sloan Digital Sky Survey \citep{2008ApJS..175..297A} supported by the Alfred P.~Sloan Foundation (a philanthropic structure set up by the one-time President of General Motors), have led this new wave, leading to deeper exposures, and more stars, than ever before. Other comparable surveys have also been carried out in the near infrared, notably the 2MASS infrared sky survey led by the University of Massachusetts \citep{2006AJ....131.1163S}.

Other very-large scale CCD or infrared sky surveys are even now coming on line, notably VST (the ESO VLT Survey Telescope), VISTA (the ESO Visible and Infrared Survey Telescope for Astronomy), and Pan-STARRS (the Panoramic Survey Telescope and Rapid Response System), while yet grander projects (notably LSST, the Large Synoptic Survey Telescope) are planned. They are located at premier high-altitude sites such as in the Atacama desert or perched in the mountain top observatories of Hawaii. The emphasis has evolved somewhat, to surveying the sky as quickly as possible in as many colour filters as technically feasible. They fall almost exclusively into the category of large-scale surveys (rather than parallax or reference-frame surveys). State-of-the-art astrometric accuracy is not their primary objective, and all have based their overall reference frame on the positional network provided by the Hipparcos Catalogue derived from the first astrometric survey from space. New challenges come as these unprecedented surveys scan the night skies, over and over, with a speed and sensitivity inconceivable only a couple of decades before.

\subsection{Nearby Stars}

The definition of the nearby stellar population figures in many areas of astronomical research, ranging from studies of star formation to the statistical occurrence of extra-solar planets. It remains, however, a difficult task to establish a complete census of stars within the immediate solar neighbourhood, even out to distances of only 10--20~parsec. 

One of the first attempts to compile a census of stars in the solar neighbourhood, largely based on trigonometric parallaxes, was Woolley's {\it `Catalogue of Stars within Twenty-Five Parsecs of the Sun'}, while a growing compilation has been maintained by the Astronomisches Rechen-Institut in Heidelberg over the last 50~years. The 1957 {\it `Katalog der Sterne  n\"aher als 20~Parsek f\"ur~1950.0'} contained 915~single stars and systems within 20~parsec.  The 1969 {\it `Catalogue of Nearby Stars'}, or CNS2, had a slightly enlarged distance limit of 22.5~parsec. CNS3 extended the census to some 1700 stars nearer than 25~parsec, while the as-yet-unpublished CNS4 incorporates data from the Hipparcos catalogue, and provides a major development in the comprehensive inventory of the solar neighbourhood up to a distance of 25~parsec from the Sun. Other compilations include Northern Arizona University `NStars Database', dating from 1998, which maintains a compilation of all stellar systems within 25~parsec, while Georgia State University's `Research Consortium on Nearby Stars' (RECONS) aims to discover and characterise `missing' stars within 10~parsec, using astrometry, photometry, and spectroscopy.

While the earliest ground-based parallax surveys were very successful in identifying nearby very bright stars, problems still persist for stars of very low intrinsic luminosity, where a complete parallax survey even out to only 10\,pc remains impossible. The advent of accurate all-sky multi-colour surveys has facilitated the direct search for nearby, low-luminosity stars. 

As Wilhelm Struve had originally suggested almost two centuries ago, surveys searching for high-proper motion stars have long been used to detect nearby candidate stars which were then added to parallax programmes, including the Hipparcos Input Catalogue in the early 1980s. Although these high-proper motion surveys imply a strong bias towards high-velocity objects, frequently part of the extended spherical `halo' component of our Galaxy's stellar population, the latest deep digital sky surveys continue to discover faint high proper motion stars, and specific attempts to determine their parallaxes are being made with the objective of completing the census of stars nearest to the Sun.

\subsection{Narrow-Field Astrometry}
\label{sec:narrow-field}

Another specialised and productive field of astrometry over the past century or more has been the study of binary and multiple stars. Many stars are born as members of a binary system (or less commonly as a triple or quadruple system), and the relative motions of their individual components, or their photocentre, has led to an enormous body of data on binary and multiple star orbits. Traditionally, long-focus telescopes with a large photographic plate scale were used. Reasonably high {\it relative\/} positional accuracy could be achieved because the atmosphere does not impose the same type of deleterious random image motion on very small angular scales (say, within 5--10~seconds of arc), as it does on larger angular scales. Accordingly, while not providing information on parallaxes, or on the celestial reference frame, this approach has provided a wealth of data on higher-order positional effects that modify relative positions on small angular scales. 

Within the last 10~years or so, this technique is being further applied to narrow-field astrometry using optical or infrared interferometers on Earth \citep[e.g.][]{1992A&A...262..353S}. Relative accuracies of order one thousandths of an second of arc or better have been achieved, while efforts are ongoing to drive these narrow-field astrometric measurements to perhaps some 10~millionths of a second of arc (as targeted by VLTI--PRIMA). Such accuracies would greatly assist in characterising the properties of the extra-solar planets now being discovered.

\section{The Move to Space}

Two thousand years of charting the stars has led us on a remarkable voyage of discovery. The Earth, as we now know, is not at all at the centre of the Universe, but a spinning body of unremarkable mass which orbits the Sun. Billions of other stars, as well as planets, interstellar gas and dust, radiation, and invisible material are bound together to form our Galaxy---a magnificent disk spiral system, prevented from collapsing by its own rotation.  Our Sun lies way out in one of the spiral arms, thirty thousand light-years from the centre. Around us the stars, at truly immense distances, move along their own eternal paths. Beyond our own island universe, the Milky Way, a seeming infinity of other galaxies recede from us at astonishing speeds, pointing their fingers backwards in time to the dawn of creation. 

Many of these advances in our understanding have accrued from a steady refinement in measuring star positions. Over the past century, improvements advanced along a very high accuracy branch for a very few stars, culminating in the compilations of parallax distances for around eight thousand stars. A medium accuracy branch for a thousand or so stars gave our very best, but still troublingly inadequate, celestial reference system. 

The lower accuracy branch developed progressively from Tycho's catalogue of a thousand stars with an accuracy of fifty seconds of arc in around 1600, Flamsteed's survey of three thousand stars to twenty seconds of arc around 1700, Lalande's fifty thousand stars at three seconds of arc around 1780, and Argelander's survey of more than three hundred thousand stars at one second of arc around the 1850s. Billion star surveys were compiled from the world's arsenal of Schmidt telescopes in the late 1900s, but despite their colossal strength in numbers, their positions were only marginally better than the celestial surveys of more than a century before.

At the dawn of the third millennium, the quality of star positions lagged far behind the progress achieved in many other areas of astronomy. Accurate distances were still only known for a few hundred nearby stars, a severe barrier to progress in understanding the physical processes within them. Accuracies from the large photographic surveys were strongly limited by the atmosphere. Proper motions were known for millions of stars, but with distortions in their systematic errors over the sky which confounded their interpretation. Distance information needed to transform them to space motions was all but lacking. 

By the second half of the twentieth century the steady advance in the accuracy of stellar positions was running headlong into a number of essentially insurmountable barriers. The biggest problem was the bending and twinkling effects of the atmosphere, condemning star images to their eternal and unpredictable wobbling dance. New thin-mirror telescope technologies have had great success in correcting effects over small angles, but all attempts to nail down large angles across the sky failed miserably. In addition, there were the tiny variations in telescope alignment as the mountain-top observatories went through their endless day and night cycles of warming and cooling. The variable flexing of telescopes under their own weight as the huge supporting structures were steered to observe different parts of the sky added other unpredictable distortions.

Yet another complication was that any telescope on Earth can observe only part of the sky at any one time: a telescope in the northern hemisphere only ever sees the northern skies. Even so, it still requires a year to elapse for the entire region to be observable by night. A grid of star positions spanning the entire sky could only be constructed from a vast spider web of thousands of geometrical triangulations from separate telescopes observing accessible portions of the sky at different times. However, between the various observations which had to be carefully patched together, all of the star images had moved by the tiny amounts which were to be probed.

Like an ancient cartographic survey of the Earth made with primitive surveying instruments, the result of centuries of effort was a map of the sky of sorts, but one which was highly distorted and unpredictably warped. At accuracies below a second of arc, it was simply unreliable. Star positions were plagued by unfathomable errors which could not be unravelled. Their space motions were, in consequence, of variable and sometimes questionable quality. More importantly, distances remained largely unknown, the tiny signatures of their minuscule parallaxes buried under a shroud of error-prone measurements imposed by the flickering atmosphere. A fundamentally new approach to measuring star positions was desperately required.

The proposal to make these delicate observations from space was the next master stroke of instrumental creativity. It  was first formally laid out in front of other scientists in the mid-1960s by 61-year old French astronomer Pierre Lacroute. Until then space science, still very much in its first flush of youth, had been somewhat the preserve of magnetospheric experts studying the region of the Earth's environment controlled by its magnetic field, discovered by Explorer--1 in 1958. X-ray astronomers, meanwhile, were eagerly following up their discovery of the first cosmic X-ray source in 1962. It seems even more remarkable in hindsight that such a specialised goal in space science should have followed so closely on the heels, within just a decade, of the first ever artificial satellite to orbit the Earth, the Soviet Union's Sputnik~1 in 1957.

Lacroute had realised that a space telescope would allow the measurement of arcs and triangulations to be made above the flickering effects of the atmosphere. Also, beyond the buckling forces of Earth's gravity, the telescope would not be sagging unpredictably as it made its cosmic census. Far from the Earth, the satellite would have an uninterrupted view of the entire sky, and the experiment could also be shielded to simulate perpetual night time. The most ingenious part of Lacroute's idea, however, was to observe in two very widely separated directions at the same time. Combining these two different sight lines into a single telescope focus, by means of a special split mirror looking out in two directions simultaneously, would give a network of wide-angle measurements spanning the whole celestial sphere in its entirety. 

The idea of making differential angular measurements was not new in itself, and indeed Friedrich Bessel's first parallax measurements had made use of a somewhat similar approach a century and a half before. The novelty, empowered by the elimination of the atmosphere, was making these differential angular measurements across very wide sweeps of the night sky. From the network of space measurements, strict trigonometric distances could be disentangled. The goal, in short, was to construct a vastly improved census of stellar parallaxes, so that their distances could be measured and their physical properties derived. The satellite concept was duly named Hipparcos, a somewhat contrived, and thereafter rarely used, contraction of `high-precision parallax collecting satellite', but also paying tribute to the ancient Greek pioneer of celestial mapping.

\begin{figure}[t]
\centering
\includegraphics[width=0.55\linewidth]{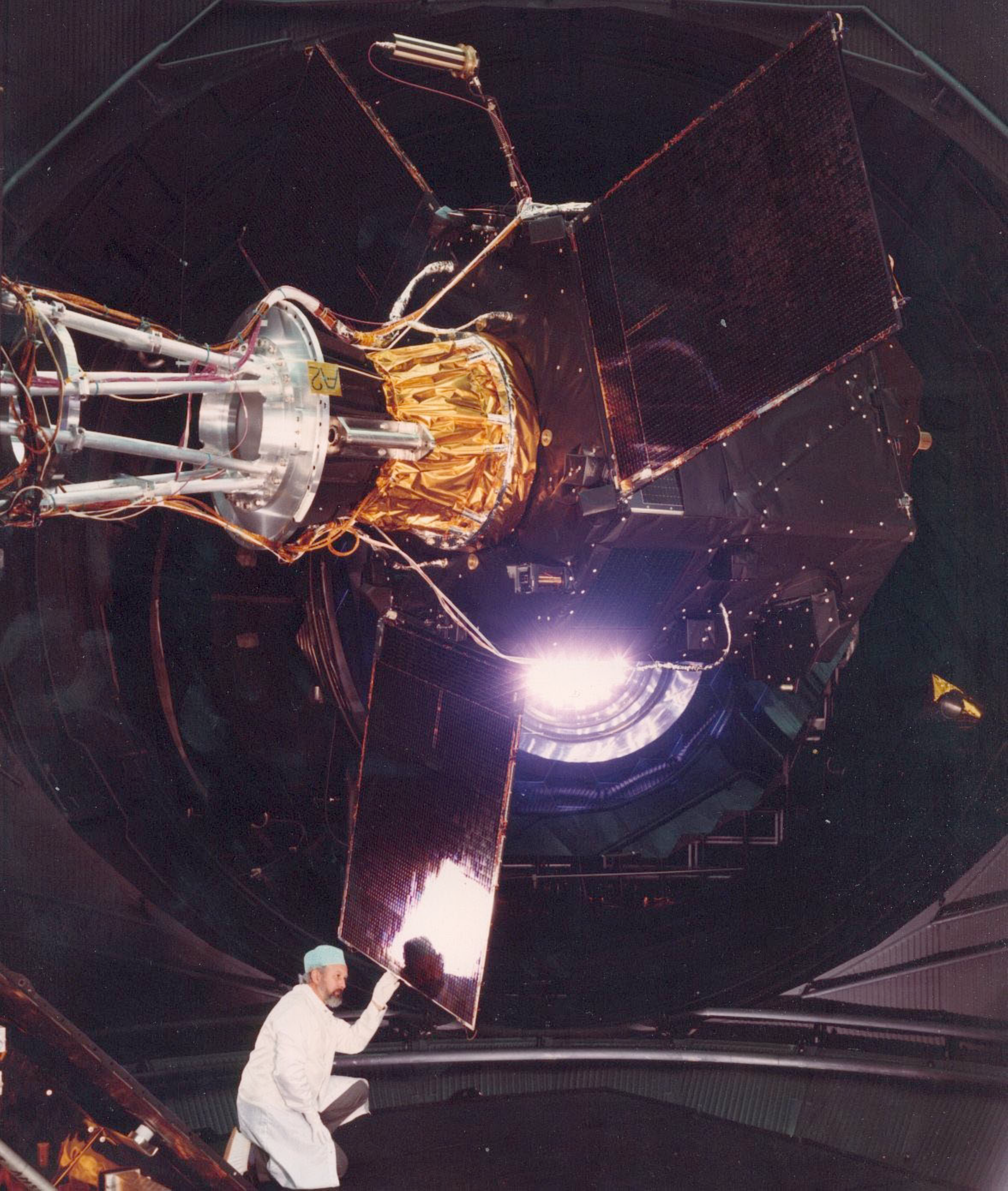}
\caption{\footnotesize Pre-launch testing of the Hipparcos satellite in the Large Solar Simulator of the European Space Agency's technical centre (ESTEC, Noordwijk, The Netherlands), February 1988.
\label{fig:satellite-lss-estec}
}
\end{figure}

A long process of lobbying, and detailed design and feasibility study, eventually led to the Hipparcos project's adoption by the European Space Agency in 1980, and the satellite's launch in 1989 (Figure~\ref{fig:satellite-lss-estec}). Particularly influential in picking up Lacroute's concept, refining its technical precepts, consolidating its mathematical foundation, and detailing its scientific objectives were the four scientific consortium leaders who dedicated much of their own careers to its successful pursuit -- Erik H{\o}g (Copenhagen), Jean Kovalevsky (Grasse), Lennart Lindegren (Lund) and Catherine Turon (Paris--Meudon). A substantial technical and scientific effort underpinned the extensive international collaboration coordinated by ESA and directed by the Hipparcos Science Team, in total comprising some 200~European scientists, 30~European industrial teams, some hundreds of engineers and managers from across the ESA member states, and an overall budget of some \textgreek{\euro}400~million (at year 2000 economic conditions). 

Publication of the Hipparcos catalogue in 1997 presented the positions, space motions, and distances of more than a hundred thousand stars, all measured with equal attention, all accurate to around one thousandth of a second of arc \citep{1997ESASP1200.....P}, comprising comprehensive astrometric \citep{1997A&A...323L..49P}, photometric \citep{1997A&A...323L..61V}, and double star data \citep{1997A&A...323L..53L}. Subsequently-published products included the Tycho~2 catalogue of 2.5~million stars \citep{2000A&A...355L..27H}, and an improvement in the astrometric quality primarily of the brightest stars \citep{2010SSRv..151..209V}.

\begin{figure}[t]
\centering
\includegraphics[width=1.0\linewidth]{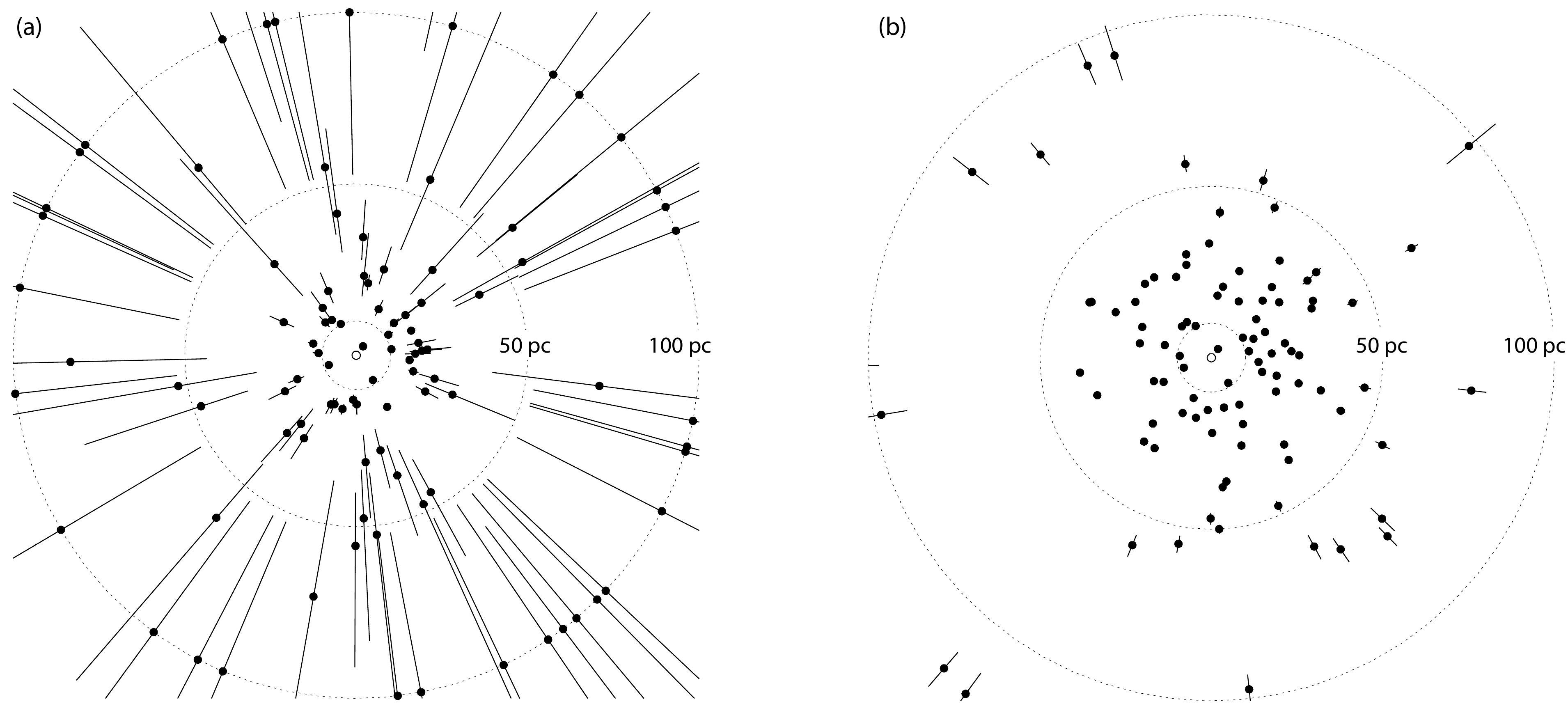}
\caption{\footnotesize Improvement in the knowledge of exoplanet host star distances by Hipparcos, as an example of the accuracy advances gained from space-based measurements in the 1990s. For the 100 brightest stars with exoplanets known from radial velocity measurements at the end of 2010 ($V<7.2$~mag), estimated distances and standard errors are shown from: (a)~the best ground-based compilation \citep{1995gcts.book.....V}, and (b)~from Hipparcos \citep{1997ESASP1200.....P}. Azimuthal coordinates correspond to right ascension, independent of declination. Distances undetermined in~(a) are arbitrarily assigned a parallax of $10\pm9$\,milli-seconds of arc. Hipparcos substantially improved both the parallax standard errors, and their systematics.
\label{fig:distances-gcstp-hipparcos}
}
\end{figure}

The Hipparcos satellite mission -- two decades of focused work by hundreds of European scientists and engineers -- provided not only the most accurate positional survey to date by far (Figure~\ref{fig:distances-gcstp-hipparcos}). Very significantly, it joined together in a single survey the most delicate work on individual stellar distances, the highest accuracy of the best reference frames, and the formidable large-scale surveys of history's great star charts. Its substantial leap in accuracy was the largest single advance in astrometry ever made in the entire history of the field, an improvement over its predecessors by a factor of fifty (Figure~\ref{fig:accuracy}), and with resulting contributions to stellar astrophysics, the distance scale, and Galactic structure and dynamics \citep{2009aaat.book.....P, 2011A&ARv..19...45P}. In the opinion of \citet{1988dyson}: {\it ``Hipparcos is the first time since Sputnik in 1957 that a major new development in space science has come from outside the United States.''} I was the ESA `project scientist' for Hipparcos for its 17-year duration (1981--1997), and further details of the organisational and sociological aspects are given in my popular account \citep{2010perryman}.

Meanwhile, also based in Earth orbit, the NASA/ESA Hubble Space Telescope, launched in 1990, has also provided narrow-field accuracies of better than one thousandth of a second of arc on a limited number of stars. As well as providing a number of key parallax determinations, the measurements made with its attitude control `Fine Guidance Sensors' have provided valuable results on a number of exoplanet systems \citep[e.g.][and references]{2010ApJ...715.1203M}. Like Hipparcos, this ambitious instrumental advance has also further validated the approach of performing high-accuracy astrometric measurements from space.

\section{The Future}

History has not come to an end.  Already by 1997, as the Hipparcos catalogue was being lodged in scientific libraries around the world, astronomers were advancing ideas for yet more ambitious experiments to map the stars from space. These included both `pointed' and `sky-scanning' instrumental approaches, amongst them the German DIVA satellite mission, NASA's Space Interferometry Mission (SIM, later SIM PlanetQuest), various initiatives from the US Naval Observatory (FAME, AMEX, OBSS, and MAPS), from Russia (OSIRIS and LIDA), and Japan (JASMINE and Nano-JASMINE). Many of these have since fallen by the wayside due to technological, cost, or political considerations, itself underlining the continued and substantial technical complexity and cost of undertaking astrometric observations from space.

The next European instrumental advance, Gaia, is at the time of writing within a year or so of launch. It will follow the same principles as Hipparcos, but with both scientific ambition and the experiment itself scaled up to reflect twenty years of progress in astronomy and technology, surpassing the Hipparcos accuracy by a factor one hundred. It will feature a larger lightweight telescope, built from the highly stable ceramic silicon carbide. Like a massive digital video camera, a carpet of CCD silicon sensors almost a square meter in area will record the millions of star images that pass across it as a new orbiting satellite once more scans the heavens. A powerful on-board processor will handle a vast cascade of image manipulations before the information stream is despatched to Earth. Its data rate will be more than a hundred times that of its predecessor. The satellite's orbit will be rather different---far from Earth, one and half million kilometers away, at the Sun--Earth Lagrange point. 

After five years of studies, and after protracted discussion and intense lobby, the European Space Agency's advisory bodies signed up to Gaia in October 2000, twenty years after a very different body of scientists did the same for Hipparcos in 1980 \citep{2001A&A...369..339P, 2012Ap&SS.tmp...68D}. It targets measurements of ten {\it millionths\/} of a second of arc for the brightest stars, a hundred times better than the pioneering results obtained from space by Hipparcos, the width of human hair viewed from a thousand kilometers. It is due for launch from Europe's space port in Kourou, French Guyana, in 2013, almost 25~years since the launch of Hipparcos. After a five year programme of scanning the skies at the start of the third millennium, its final harvest will be in scientific hands in 2020. 

This next leap in ambition promises a scientific harvest which dwarfs that of Hipparcos. Its colossal survey of more than a thousand million stars will provide a defining census of around one per cent of our Galaxy's entire stellar population, pin-pointing them in space right across its vast expanses. Unimaginable numbers of stellar motions will reveal many more details of the vastly complex motions at play within our Galaxy. It will provide insights ranging from new tests of general relativity to stringent limits on the variation of fundamental physical constants. Planets circling other stars will appear in their thousands from their tiny wobbling motions, identifying candidate systems for the burgeoning discipline of exo-biology. Tens of thousands of asteroids will be measured. Objects which may approach Earth in the coming century will have their trajectories plotted, and if their projected orbit suggests a collision course, we will perhaps have time enough to see whether mankind's innate resourcefulness can do something to avert an impending and potentially calamitous impact. 

Perhaps, in a decade or two from now, some ingenious scientists and space engineers will have figured out how to build a satellite to measure at the levels of a thousand times better than Gaia, at the billionth of a second of arc.  At that point, distances out across the vast uncharted cosmological expanses of the Universe could be measured directly. 

For now, such a possibility remains firmly in the realms of science fiction. Indeed, as Danish authority Erik H{\o}g has written after his lifelong contributions to the field, and based on his recent studies of its historical and technical development: {\it ``The Gaia astrometric survey of a thousand million stars cannot be surpassed in completeness and accuracy within the next forty or fifty years.''}  History is littered with erroneous predictions, so many self-proclaimed seers consistently failing to anticipate the accelerating pace of change. It would take a brave person to wager a significant sum either way, but my tendency would be to side with Erik H{\o}g!

\section*{Acknowledgments}

In preparing this review, my text on the early history of astrometry draws on the cited works of David Goodman \& Colin Russell (1991), Michael Hoskin (1997), and Allan Chapman (1990). The latter provides numerous further references to developments of critical angular measurement in astronomy over the period 1500--1850. The basis of this account, including a number of the figures, appeared in my popular book describing the Hipparcos project {\it The Making of History's Greatest Star Map} \citep{2010perryman}. I am grateful to my many colleagues involved in the Hipparcos project who, over many years, stimulated and advanced my own involvement in astrometry, as well as in its history. 

I thank Virginia Trimble for her invitation to prepare this article, and for her valued guidance and suggestions during its preparation. I would like to stress that my coverage of the developments in astrometry over the past century is intentionally somewhat superficial and is certainly far from complete, being targeted more for those interested in an overview of the subject in its broadest outlines, rather than for those with a deeper focus on astronomy.


\bibliographystyle{apalike}
\footnotesize
\bibsep 2pt
\bibliography{history}

\end{document}